\documentclass[cmp,american]{svjour}
\usepackage{amsmath}
\usepackage{graphics}
\usepackage{amssymb}

\providecommand{\tabularnewline}{\\}

\usepackage{babel}

\begin{document}

\title{Exotic smooth $\mathbb{R}^{4}$, noncommutative algebras and quantization}

\author{Torsten Asselmeyer-Maluga\inst{1} \and Jerzy Kr\'ol\inst{2}}
\institute{German Aero space center, Rutherfordstr. 2, 12489 Berlin, torsten.asselmeyer-maluga@dlr.de%
\and 
University of Silesia, ul. Uniwesytecka 4, 40-007 Katowice, iriking@wp.pl %
}
\date{Received: date / Accepted: date}
\communicated{name}
\maketitle
\begin{abstract}
The paper shows deep connections between exotic smoothings of small
$\mathbb{R}^{4}$, noncommutative algebras of foliations and quantization.
At first, based on the close relation of foliations and noncommutative
$C^{\star}$-algebras we show that cyclic cohomology invariants characterize
some small exotic $\mathbb{R}^{4}$. Certain exotic smooth $\mathbb{R}^{4}$'s
define a generalized embedding into a space which is $K$-theoretic
equivalent to a noncommutative Banach algebra. 

Furthermore, we show that a factor $I\! I\! I$ von Neumann algebra
is naturally related with nonstandard smoothing of a small $\mathbb{R}^{4}$
and conjecture that this factor is the unique hyperfinite factor $I\! I\! I_{1}$.
We also show how an exotic smoothing of a small $\mathbb{R}^{4}$
is related to the Drinfeld-Turaev (deformation) quantization of the
Poisson algebra \emph{$(X(S,SL(2,\mathbb{C}),\left\{ \,,\,\right\} )$}
of complex functions on the space of flat connections \emph{$X(S,SL(2,\mathbb{C})$}
over a surface $S$, and that the result of this quantization is the
skein algebra $(K_{t}(S),[\,,\,])$ for the deformation parameter
$t=\exp(h/4)$. This skein algebra is retrieved as a $I\! I_{1}$
factor of horocycle flows which is Morita equivalent to the $I\! I_{\infty}$
factor von Neumann algebra which in turn determines the unique factor
$I\! I\! I_{1}$ as crossed product $I\! I_{\infty}\rtimes_{\theta}\mathbb{R}_{+}^{*}$.
Moreover, the structure of Casson handles determine the factor $I\! I_{1}$
algebra too. Thus, the quantization of the Poisson algebra of closed
circles in a leaf of the codimension 1 foliation of $S^{3}$ gives
rise to the factor $I\! I\! I_{1}$ associated with exotic smoothness
of $\mathbb{R}^{4}$. 

Finally, the approach to quantization via exotic 4-smoothness is considered
as a fundamental question in dimension 4 and compared with the topos
approach to quantum theories. 
\end{abstract}
\tableofcontents{}

\section{Introduction}

Even though non-standard smooth $\mathbb{R}^{4}$'s exist as a 4-dimensional
smooth manifolds, one still misses a direct coordinate-like presentation
which would allow the usage of global, exotic smooth functions. Such
functions are smooth in the exotic smoothness structure but fail to
be differentiable in a standard way determined by the topological
product of axes. On the way toward uncovering peculiarities of exotic
$\mathbb{R}^{4}$'s we obtained some unexpected connections of these
with quantum theories and string theory formalism, see \cite{AsselmeyerKrol2009,AsselmeyerKrol2009a}.
It seems however that any complete knowledge of the connection especially
to quantum theories is still missing. The whole complex of problems
is strongly connected to a future theory of quantum gravity where
one has to include exotic smoothness structures in the formal functional
integration over the space of metrics or connections. The presented
paper aims to fill this gap and shed light on the above-mentioned
connection with quantum theories. The relation with string theory
will be addressed in our forthcoming paper. At first we will show
how exotic smooth $\mathbb{R}^{4}$'s are related with certain noncommutative
Banach and $C^{\star}$-algebras. Then we will discuss the main result
of the paper: the explanation how the factor $I\! I\! I$ von Neumann
algebra corresponds to small exotic $\mathbb{R}^{4}$ and the explicit
quantization procedure of the Poisson algebra of loops on surfaces
driven by a small exotic $\mathbb{R}^{4}$. Moreover, the geometric
content of the quantization and the structure of this factor $I\! I\! I$
is carefully worked out. We observe and discuss the possible relation
of this kind of quantization driven by 4-exotics, with the topos approach
to (quantum) theories of physics. 

We have observed in our previous paper \cite{AsselmeyerKrol2009},
that the exoticness of some small exotic $\mathbb{R}^{4}$'s is localized
on some compact submanifolds and depend on the embedding of the submanifolds.
Or, a small exotic structures of the $\mathbb{R}^{4}$ is determined
by the so-called Akbulut cork (a 4-dimensional compact contractible
submanifold of $\mathbb{R}^{4}$ with a boundary), and its embedding
given by an attached Casson handle. The boundary of the cork is a
homology 3-sphere containing a 3-sphere $S^{3}$ such that the codimension-1
foliations are determined by the foliations of $S^{3}$. We have explained
this important point in details in our previous papers \cite{AsselmeyerKrol2009,AsselmeyerKrol2009a}.
Following this topological situation we want to localize exotica on
a compact submanifold of $\mathbb{R}^{4}$ such that we are able to
relate the topological data, like foliations and Casson handles, with
some other structures on the submanifold. The changes of the structures
were grasped already by techniques from conformal field theory and
WZW models and gerbes on groupoids and can be related with string
theory and correspond to the changes of exotic smoothness on $\mathbb{R}^{4}$
(see \cite{AsselmeyerKrol2009,AsselmeyerKrol2009a}). We follow this
philosophy further in this paper. Especially we will find:

\begin{itemize}
\item a compact submanifold of $\mathbb{R}^{4}$ such that its $K$-theory
is deformed towards the $K$-theory of some noncommutative algebra
when changing the standard smoothness on $\mathbb{R}^{4}$ toward
exotic one. This compact submanifold is again the $S^{3}$ part of
the boundary of the Akbulut cork of exotic $\mathbb{R}^{4}$, i.e.
$S^{3}\subset\mathbb{R}^{4}$. We describe the noncommutative Banach
algebra whose $K$-theory is the deformed $K$-theory of the commutative
algebra $C(S^{3})$, $S^{3}\subset\mathbb{R}^{4}$ when the smoothness
is changed from standard to exotic. We say that such an exotic $\mathbb{R}^{4}$
contains an embedded $S^{3}$ or that exotic $\mathbb{R}^{4}$ deforms
the $K$-theory of the 3-sphere. This is the content of section \ref{sub:Exotic-and-noncommutative algebra}
and theorem \ref{Th:Exotics-Banach-algebras}. 
\item Next, given a codimension-1 foliation $(S^{3},F)$ of $S^{3}$ where
$F$ is an integrable subbundle $F\subset TS^{3}$, one can associate
a $C^{\star}$- algebra $C(S^{3},F)$ to the leaf space $S^{3}/F$
of the foliation. Hurder and Katok \cite{HurKat:84} showed that every
$C^{*}$algebra of a foliation with non-trivial Godbillon-Vey invariant
contains a factor $I\! I\! I$ subalgebra. Based on the relation of
codimension-1 foliations of $S^{3}$ and small exotic smooth structures
on $\mathbb{R}^{4}$, as established in \cite{AsselmeyerKrol2009},
one has a factor $I\! I\! I$ algebra corresponding to the exotic
$\mathbb{R}^{4}$'s. In subsection \ref{sub:Leaf-space-and} we construct
the $C^{\star}$- algebra $C(M,F)$ of the foliation with a nontrivial
Godbillon-Vey invariant and relate the exotic $\mathbb{R}^{4}$ to
the factor $I\! I\! I$ von Neumann algebra. Furthermore we conjecture
that the factor $I\! I\! I$ is in fact the unique hyperfinite factor
$I\! I\! I_{1}$. In subsection \ref{sub:Cyclic-cohomology-and} we
relate the smoothing of $\mathbb{R}^{4}$ with the cyclic cohomology
invariants of the $C^{\star}$- algebra $C(M,F)$. This is done via
the KK theory of Kasparov , allowing the definition of the K theory
of the leaf space of the foliation, $K(V/F)$, and the analytic assembly
map $\mu$ of Connes and Baum using finally the Connes pairing between
K theory and cyclic cohomology.
\end{itemize}
In section \ref{sec:Quantization} we turn to the quantization procedure
as related to nonstandard smoothings of $\mathbb{R}^{4}$. Based on
the dictionary between operator algebra and foliations one has the
corresponding relation of small exotic $\mathbb{R}^{4}$'s and operator
algebras. This is a noncommutative $C^{\star}$ algebra which can
be seen as the algebra of quantum observables of some theory. 

\begin{itemize}
\item First, in subsection \ref{sub:factor-III-case} we recognized the
algebra as the hyperfinite factor $I\! I\! I_{1}$ von Neumann algebra.
From Tomita-Takesaki theory it follows that any factor $I\! I\! I$
algebra $M$ decomposes as a crossed product into $M=N\rtimes_{\theta}\mathbb{R}_{+}^{*}$
where $N$ is a factor $I\! I_{\infty}$. Via Connes procedure one
can relate the factor $I\! I\! I$ foliation to a factor $I\! I$
foliation. Then we obtain a foliation of the horocycle flow on the
unit tangent bundle over some genus $g$ surface which determines
the factor $I\! I_{\infty}$. This foliation is in fact determined
by the horocycles which are closed circles.
\item Next we are looking for a classical algebraic structure which would
give the above mentioned noncommutative algebra of observables as
a result of quantization. The classical structure is recovered by
the idempotent of the $C^{\star}$ algebra and has the structure of
a Poisson algebra. The idempotents were already constructed in subsection
\ref{sub:smooth-holonomy-groupoid} as closed curves in the leaf of
the foliation of $S^{3}$. As noted by Turaev \cite{Turaev1991},
closed curves in a surface induces a Poisson algebra: Given a surface
$S$ let $X(S,G)$ be the space of flat connections on $G=SL(2,\mathbb{C})$
bundles on $S$; this space carries a Poisson structure as is shown
in subsection \ref{sub:The-observable-algebra}. The complex functions
on $X(S,SL(2,\mathbb{C}))$ can be considered as the algebra of classical
observables forming the Poisson algebra $(X(S,SL(2,\mathbb{C})),\left\{ \,,\,\right\} )$. 
\item Next in the subsection \ref{sub:Drinfeld-Turaev-Quantization} we
find a quantization procedure of the above Poisson algebra which is
the Drinfeld-Turaev deformation quantization. It is shown that the
result of this quantization is the skein algebra $(K_{t}(S),[\,,\,])$
for the deformation parameter $t=\exp(h/4)$ ($t=-1$ corresponds
to the commutative Poisson structure on $(X(S,SL(2,\mathbb{C})),\{\,,\,\})$). 
\item This skein algebra is directly related to the factor $I\! I\! I_{1}$
von Neumann algebra derived from the foliation of $S^{3}$. In fact
the skein algebra is constructed in subsection \ref{sub:Temperley-Lieb-algebra-foliation}
as the factor $I\! I_{1}$ algebra Morita equivalent to the factor
$I\! I_{\infty}$ which in turn determines the factor $I\! I\! I_{1}$
of the foliation. 
\item Next, in subsection \ref{sub:Casson-handle-and} we show that at the
level of 4-exotic smooth structures main building blocks of these,
i.e. Casson handles, determine the factor $I\! I_{1}$ algebras. This
was already shown in \cite{Asselmeyer2007,AssRos:05}. 
\item Finally, in subsection \ref{sub:Topoi} we argue that the appearance
of the factor $I\! I\! I_{1}$ and quantization in context of small
exotic $\mathbb{R}^{4}$'s can be lifted to a general and fundamental
tool applicable for local relativistic algebraic quantum field theories
in dimension 4. This is quite similar to the topos approach to (quantum)
theories as has been recently actively developed (see e.g. \cite{Isham:08,Krol2006a,Landsman:07}).
We show that under some natural suppositions regarding the role of
the factor $I\! I\! I_{1}$ and the quantization of the Poisson algebra
$(X(S,SL(2,\mathbb{C})),\left\{ \,,\,\right\} )$ of complex functions
on $X(S,SL(2,\mathbb{C})$, in field theories, our 4-exotics approach
can be seen as a way how to reformulate quantum theories and make
them classical. The price to pay is the change of smoothing of $\mathbb{R}^{4}$
and this works in dimension 4. In the topos approach one is changing
logic and set theory into intuitionistic ones valid in topoi and this
is quite universal procedure. However, we think that the specific
choice of dimension 4 by our 4-exotics formalism can be particularly
well suited to the task of quantization of gravity in dimension 4
(cf. \cite{Krol:04a,Krol:2005,AssRos:05}). 
\end{itemize}
Then we have a closed circle: the exotic $\mathbb{R}^{4}$ determines
a codimension-1 foliation of a 3-sphere which produces a factor $I\! I\! I_{1}$used
in the algebraic quantum field theory (see \cite{Borchers2000}) as
vacuum. In \cite{AsselmeyerKrol2009} we constructed a relation between
the leaf space of the foliation and the disks in the Casson handle.
Thus the complicated structure of the leaf space is directly related
to the complexity of the Casson handle making them to a noncommutative
space.

\section{Exotic $\mathbb{R}^{4}$ and codimension-1 foliation\label{sec:Exotic-R4-foliation}}

Here we present the main line of argumentation in our previous paper
\cite{AsselmeyerKrol2009}:

\begin{enumerate}
\item In Bizacas exotic $\mathbb{R}^{4}$ one starts with the neighborhood
$N(A)$ of the Akbulut cork $A$ in the K3 surface $M$. The  exotic
$\mathbb{R}^{4}$ is the interior of $N(A)$.
\item This neighborhood $N(A)$ decomposes into $A$ and a Casson handle
representing the non-trivial involution of the cork.
\item From the Casson handle we construct a grope containing Alexanders
horned sphere.
\item Akbuluts construction gives a non-trivial involution, i.e. the double
of that construction is the identity map.
\item From the grope we get a polygon in the hyperbolic space $\mathbb{H}^{2}$.
\item This polygon defines a codimension-1 foliation of the 3-sphere inside
of the exotic $\mathbb{R}^{4}$ with an wildly embedded 2-sphere,
Alexanders horned sphere. This foliation agrees
with the corresponding foliation of the homology 3-sphere $\partial A$.
This codimension-1 foliations of $\partial A$ is partly classified
by the Godbillon-Vey class lying in $H^{3}(\partial A,\mathbb{R})$
which is isomorphic to $H^{3}(S^{3},\mathbb{R})$.
\item Finally we get a relation between codimension-1 foliations of the
3-sphere and exotic $\mathbb{R}^{4}$.
\end{enumerate}
This relation is very strict, i.e. if we change the Casson handle
then we must change the polygon. But that changes the foliation and
vice verse. Finally we obtained the result:\\
\emph{The exotic $\mathbb{R}^{4}$ (of Bizaca) is determined by the
codimension-1 foliations with non-vanishing Godbillon-Vey class in
$H^{3}(S^{3},\mathbb{R}^{3})$ of a 3-sphere seen as submanifold $S^{3}\subset\mathbb{R}^{4}$.}

\section{Exotic $\mathbb{R}^{4}$ and operator algebras \label{sec:Gerbes-and-exotic}}

In our previous paper \cite{AsselmeyerKrol2009} we uncover a relation
between an exotic (small) $\mathbb{R}^{4}$ and non-cobordant codimension-1
foliations of the $S^{3}$ classified by Godbillon-Vey class as element
of the cohomology group $H^{3}(S^{3},\mathbb{R})$. By using the $S^{1}$-gerbes
it was possible to interpret the integer elements $H^{3}(S^{3},\mathbb{Z})$
as characteristic class of a $S^{1}$-gerbe over $S^{3}$. We also
discuss a possible deformation to include the full real case as well.
Here we will use the idea to relate an operator algebra to the foliation.
Then the invariants of the operator algebra will reflect the invariants
of the foliation.

\subsection{Twisted $K$-theory and algebraic $K$-theory\label{sub:Twisted-K-theory-algebraic} }

It is known that ordinary $K$-theory of a (compact) manifold $M$
is the algebraic $K$-theory of the (commutative) algebra of continuous,
complex valued functions on $M$, $C(M)$ (\cite{Connes94}, Chap.
2, Sec.1). Then twisted $K$-theory of $M$ (w.r.t. the twisting $[H]\in H^{3}(M,\mathbb{Z})$)
can be similarly determined as an algebraic $K$-theory of some generalized
algebra. In fact this algebra must be a noncommutative (non-unital)
Banach algebra which we are going to describe now \cite{AtiyahSegal2004}. 

To describe twisted $K$-theory on a manifold $M$ one can follow
the idea of ordinary non-twisted $K$-theory to construct a (representation)
space $A$ such that $K^{0}(M)=[M,A]$ where $[M,A]$ is the space
of homotopy classes of continuous maps. As shown by Atiyah \cite{Atiyah1967},
the representation space for the $K$-theory of $M$ is the topological
space $Fred({\cal H})$ of Fredholm operators in an infinite dimensional
Hilbert space ${\cal H}$, with the norm topology . Thus 

\[
K^{0}(M)=[M,Fred({\cal H})]\:.\]

This is based on the fact that a family of deformations of Fredholm
operators parametrized by $M$ whose kernel and co-kernel have locally
constant dimensions, are given by vector bundles on $M$. The formal
difference of these vector bundles determines the element of $K^{0}(M)$.

The twisting of this $K$-theory can be performed by the elements
of $H^{3}(M,\mathbb{Z})$. One of the interpretations of these integral
3-rd cohomology classes is by the projective, infinite dimensional
bundles on $M$, namely given a class $[H]\in H^{3}(M,\mathbb{Z})$
we can represent it by a projective $PU({\cal H})$ bundle $Y$ whose
class is $[H]$. Here $U({\cal H})$ is the group of unitary operators
on ${\cal H}$. Let us see briefly how it is possible \cite{AtiyahSegal2004,AsselmeyerKrol2009a}. 

The classifying space of the third cohomology group of $M$ is the
Eilenberg-MacLane space $K(\mathbb{Z},3)$. The projective unitary
group on ${\cal H}$, $PU({\cal H})=U({\cal H})/U(1)$, can now be
determined. A model for $K(\mathbb{Z},3)$ is the classifying space
of $PU({\cal H})$, i.e., $K(\mathbb{Z},3)=BPU({\cal H})$. This means
that $H^{3}(M,\mathbb{Z})=[M,K(\mathbb{Z},3)]=[M,BPU({\cal H})]$
and the realization of $H^{3}(M,\mathbb{Z})$ is as follows:

\emph{Isomorphism classes of principal $PU({\cal H})$ bundles over
$M$ correspond one to one to the classes from $H^{3}(M,\mathbb{Z})$.}

Now let us associate to a class $[H]$ (torsion or not) a $PU({\cal H})$
bundle $Y$ representing the class. Let $Fred({\cal H})$ acts on
\emph{$PU({\cal H})$} by conjugations \emph{. }We can form an associated
bundle 

\[
Y(Fred)=Y\otimes_{PU({\cal H})}Fred({\cal H})\]
Let $[M,Y(Fred)]$ denote the space of all homotopy classes of sections
of the bundle $Y(Fred)$. Then one can define the twisted $K$-theory: 

\begin{definition}

The twisted by $[H]\in H^{3}(M,\mathbb{Z})$ $K$-theory of $M$,
i.e. $K(M,[H])$ is given by the homotopy classes $[M,Y(Fred)]$ of
the sections of $Y(Fred)$, i.e. 

\begin{equation}
K(M,[H])=[M,Y(Fred)]\label{eq:K(M,[H])}\end{equation}

\end{definition} 

Now we are searching for bundles of algebras on $M$ whose sections
would be an algebra such that its $K$-theory is the $K(M,[H])$.
One natural candidate is the bundle $End(P)$ of endomorphisms of
the projective $PU({\cal H})$-bundle $P$ representing the class
$[H]\in H^{3}(M,\mathbb{Z})$. However, this choice gives rise to
the trivial $K$-theory \cite{AtiyahSegal2004}. Instead one should
take the algebra ${\cal K}$ of compact operators in ${\cal H}$ with
the norm topology. This is a noncommutative Banach algebra without
unity. Therefore, we attach to the projective $PU({\cal H})$ bundle
$P$, representing $[H]\in H^{3}(M,\mathbb{Z})$, the bundle ${\cal K}_{P}$
of non-unital algebras. The fiber of the bundle ${\cal K}_{P}$ at
$x\in M$ is the algebra of compact operators acting on $P_{x}$.
The algebra of sections of the bundle ${\cal K}_{P}$ is an noncommutative
Banach non-unital algebra $\Gamma{\cal K}_{P}$. Now one obtains the
result (\cite{AtiyahSegal2004}, Definition 3.4 and Theorem 3.2 of
\cite{Segal2001}):

\begin{theorem}\label{Th:NoncomBanachAlgebra}

Let $M$ be a compact manifold and $[H]\in H^{3}(M,\mathbb{Z})$,
the group $K(M,[H])$ is canonically isomorphic to the algebraic $K$-theory
of the noncommutative non-unital Banach algebra $\Gamma{\cal K}_{P}$.

\end{theorem}

\subsection{Exotic $\mathbb{R}^{4}$ and noncommutative Banach algebras\label{sub:Exotic-and-noncommutative algebra}}

The interpretation of twisted $K$-theory for a manifold $M$ as the
algebraic $K$-theory of some noncommutative Banach algebra can be
seen as deformation of a space towards the noncommutative space using
the twisting of $K$-theory. We are interested in the following special
smooth submanifolds of $M$:

\begin{definition}\label{Def-Gen-Embedding}\emph{ }We say that a
compact submanifold $L\subset M$ of a manifold $M$ is \emph{embedded
in $M$} \emph{as a noncommutative subspace }or\emph{ embedded in
a generalized smooth sense} when the following two conditions hold

\begin{enumerate}
\item $L$ is embedded in $M$ in ordinary sense, 
\item when a smooth structure on $M$ is changed, the $K$ -theory of $L$
is deformed toward the algebraic $K$ - theory of some noncommutative
Banach algebra.
\end{enumerate}
\end{definition}

From \cite{AsselmeyerKrol2009a}, Theorem 2, we know that small exotic
$\mathbb{R}^{4}$'s (corresponding to the integral 3-rd cohomologies
of $S^{3}$) deform $K$-theory of $S^{3}$ toward twisted $K$-theory.
Twisted $K$-theory of the compact manifold $S^{3}$ by $[H]\in H^{3}(S^{3},\mathbb{Z})$,
is in turn the $K$-theory of the noncommutative Banach algebra $\Gamma{\cal K}_{P}$
as we explained in the previous subsection.

In our case of $S^{3}$ we have the twisted $K$-theory with twist
$\tau=k[\,]\in H^{3}(S^{3},\mathbb{Z})$:

\begin{equation}
K^{\tau+n}(S^{3})=\begin{cases}
\begin{array}{c}
0,n=0\\
\mathbb{Z}/k,n=1\end{array}\end{cases}\end{equation}
which can be interpreted as the algebraic $K$-theory of the noncommutative
algebra. Finally we can formulate:

\begin{theorem}\label{Th:Exotics-Banach-algebras}

Small exotic $\mathbb{R}^{4}$, corresponding to the integral 3-rd
cohomologies of $S^{3}$, deform the $K$-theory of $S^{3}$ toward
the $K$-theory of some noncommutative algebra. Thus some small non-standard
smooth structures of $\mathbb{R}^{4}$ deform embedded $S^{3}$ towards
a noncommutative space. 

\end{theorem}

The deformation is performed by using the commutative algebra of complex-valued
continuous functions on $S^{3}$ and changing it to the noncommutative
non-unital Banach algebra of sections of the bundle ${\cal K}_{P}$
on $S^{3}$:

\[
\delta_{P}:C(S^{3})\to\Gamma{\cal K}_{P}\,.\]
Here $P$ is the projective bundle on $S^{3}$ whose Dixmier-Douady
class is $[H]\in H^{3}(S^{3},\mathbb{Z})$ and the deformed exotic
smooth $\mathbb{R}^{4}$ is determined by the same class (see Subsec.
\ref{sub:Twisted-K-theory-algebraic}). We see that the exotic smoothness
of $\mathbb{R}^{4}$ is \emph{localized} on $S^{3}$ making it a noncommutative
space, while the standard smooth $\mathbb{R}^{4}$ corresponds rather
to the ordinary, non-twisted space $S^{3}\subset\mathbb{R}^{4}$ and
the commutative algebra $C(S^{3})$.

The above case of small exotic $\mathbb{R}^{4}$ and embedded $S^{3}$
in it, is the example of the generalized embedding in a sense of our
definition \ref{Def-Gen-Embedding} and this is the only known example
to us of this phenomenon. In fact, any other $\mathbb{R}^{n}$, $n\neq4$
excludes any generalized smooth embedding. There is a possibility
to be more explicit in the description of this generalized embeddings,
namely by generalized Hitchin's structures on $S^{3}$ and their relation
to exotics \cite{AsselmeyerKrol2009}. However we do not address this
issue here.

\subsection{Leaf space and factor $I\! I\! I$ $C^{*}$-algebras\label{sub:Leaf-space-and}}

Given a foliation $(M,F)$ of a manifold $M$, i.e. an integrable
subbundle $F\subset TM$ of the tangent bundle $TM$. The leaves $L$
of the foliation $(M,F)$ are the maximal connected submanifolds $L\subset M$
with $T_{x}L=F_{x}\:\forall x\in L$. We denote with $M/F$ the set
of leaves or the leaf space. Now one can associate to the leaf space
$M/F$ a $C^{*}$algebra $C(M,F)$ by using the smooth holonomy groupoid
$G$ of the foliation. For a codimension-1 foliation there is the
Godbillon-Vey invariant \cite{GodVey:71} as element of $H^{3}(M,\mathbb{R})$.
Hurder and Katok \cite{HurKat:84} showed that the $C^{*}$algebra
of a foliation with non-trivial Godbillon-Vey invariant contains a
factor $I\! I\! I$ subalgebra. In the following we will construct
this $C^{*}$algebra and discuss the factor $I\! I\! I$ case.

\subsubsection{The smooth holonomy groupoid and its $C^{*}$algebra\label{sub:smooth-holonomy-groupoid}}

Let $(M,F)$ be a foliated manifold. Now we shall construct a von
Neumann algebra $W(M,F)$ canonically associated to $(M,F)$ and depending
only on the Lebesgue measure class on the space $X=M/F$ of leaves
of the foliation. The classical point of view, $L^{\infty}(X)$, will
only give the center $Z(W)$ of $W$. According to Connes \cite{Connes94},
we assign to each leaf $\ell\in X$ the canonical Hilbert space of
square-integrable half-densities $L^{2}(\ell)$. This assignment,
i.e. a measurable map, is called a random operator forming a von Neumann
$W(M,F)$. The explicit construction of this algebra can be found
in \cite{Connes1984}. Here we remark that $W(M,F)$ is also a noncommutative
Banach algebra which is used above. Alternatively we can construct
$W(M,F)$ as the compact endomorphisms of modules over the $C^{*}$
algebra $C^{*}(M,F)$ of the foliation $(M,F)$ also known as holonomy
algebra. From the point of view of K theory, both algebras $W(M,F)$
and $C^{*}(M,F)$ are Morita-equivalent to each other leading to the
same $K$ groups. In the following we will construct the algebra $C^{*}(M,F)$
by using the holonomy groupoid of the foliation.

Given a leaf $\ell$ of $(M,F)$ and two points $x,y\in\ell$ of this
leaf, any simple path $\gamma$ from $x$ to $y$ on the leaf $\ell$
uniquely determines a germ $h(\gamma)$ of a diffeomorphism from a
transverse neighborhood of $x$ to a transverse neighborhood of $y$.
The germ of diffeomorphism $h(\gamma)$ thus obtained only depends
upon the homotopy class of $\gamma$ in the fundamental groupoid of
the leaf $\ell$, and is called the holonomy of the path $\gamma$.
The holonomy groupoid of a leaf $\ell$ is the quotient of its fundamental
groupoid by the equivalence relation which identifies two paths $\gamma$
and $\gamma'$ from $x$ to $y$ (both in $\ell$) iff $h(\gamma)=h(\gamma')$.
The holonomy covering $\tilde{\ell}$ of a leaf is the covering of
$\ell$ associated to the normal subgroup of its fundamental group
$\pi_{1}(\ell)$ given by paths with trivial holonomy. The holonomy
groupoid of the foliation is the union $G$ of the holonomy groupoids
of its leaves. 

Recall a groupoid ${\tt G}$ is a category where every morphism is
invertible. Let $G_{0}$ be a set of objects and $G_{1}$ the set
of morphisms of ${\tt G}$, then the structure maps of ${\tt G}$
reads as:

\begin{equation}
G_{1}\,_{t}\times_{s}G_{1}\overset{m}{\rightarrow}G_{1}\overset{i}{\rightarrow}G_{1}\overset{s}{\underset{t}{\rightrightarrows}}G_{0}\overset{e}{\rightarrow}G_{1}\label{eq:def-groupoid}\end{equation}
where $m$ is the composition of the composable two morphisms (target
of the first is the source of the second), $i$ is the inversion of
an arrow, $s,\, t$ the source and target maps respectively, $e$
assigns the identity to every object. We assume that $G_{0,1}$ are
smooth manifolds and all structure maps are smooth too. We require
that the $s,\, t$ maps are submersions, thus $G_{1}\,_{t}\times_{s}G_{1}$
is a manifold as well. These groupoids are called \emph{smooth} groupoids.

Given an element $\gamma$ of $G$, we denote by $x=s(\gamma)$ the
origin of the path $\gamma$ and its endpoint $y=t(\gamma)$ with
the range and source maps $t,s$. An element $\gamma$ of $G$ is
thus given by two points $x=s(\gamma)$ and $y=r(\gamma)$ of $M$
together with an equivalence class of smooth paths: the $\gamma(t)$,
$t\in[0,1]$with $\gamma(0)=x$ and $\gamma(1)=y$, tangent to the
bundle $F$ (i.e. with $\frac{d}{dt}\gamma(t)\in F_{\gamma(t)}$,
$\forall t\in\mathbb{R}$) identifying $\gamma_{1}$ and $\gamma_{2}$
as equivalent iff the holonomy of the path $\gamma_{2}\circ\gamma_{1}^{-1}$
at the point $x$ is the identity. The graph $G$ has an obvious composition
law. For $\gamma,\gamma'\in G$ , the composition $\gamma\circ\gamma'$
makes sense if $s(\gamma)=t(\gamma)$. The groupoid $G$ is by construction
a (not necessarily Hausdorff) manifold of dimension $\dim G=\dim V+\dim F$.
We state that $G$ is a smooth groupoid, the \emph{smooth holonomy
groupoid}.

Then the $C^{*}$algebra $C_{r}^{*}(M,F)$ of the foliation $(M,F)$
is the $C^{*}$algebra $C_{r}^{*}(G)$ of the smooth holonomy groupoid
$G$. For completeness we will present the explicit construction (see
\cite{Connes94} sec. II.8). The basic elements of $C_{r}^{*}(M,F)$)
are smooth half-densities with compact supports on $G$, $f\in C_{c}^{\infty}(G,\Omega^{1/2})$,
where $\Omega_{\gamma}^{1/2}$ for $\gamma\in G$ is the one-dimensional
complex vector space $\Omega_{x}^{1/2}\otimes\Omega_{y}^{1/2}$, where
$s(\gamma)=x,t(\gamma)=y$, and $\Omega_{x}^{1/2}$ is the one-dimensional
complex vector space of maps from the exterior power $\Lambda^{k}F_{x}$
,$k=\dim F$, to $\mathbb{C}$ such that \[
\rho(\lambda\nu)=|\lambda|^{1/2}\rho(\nu)\qquad\forall\nu\in\Lambda^{k}F_{x},\lambda\in\mathbb{R}\:.\]
For $f,g\in C_{c}^{\infty}(G,\Omega^{1/2})$, the convolution product
$f*g$ is given by the equality\[
(f*g)(\gamma)=\intop_{\gamma_{1}\circ\gamma_{2}=\gamma}f(\gamma_{1})g(\gamma_{2})\]
Then we define via $f^{*}(\gamma)=\overline{f(\gamma^{-1})}$ a $*$operation
making $C_{c}^{\infty}(G,\Omega^{1/2})$ into a $*$algebra. For each
leaf $L$ of $(M,F)$ one has a natural representation of $C_{c}^{\infty}(G,\Omega^{1/2})$
on the $L^{2}$ space of the holonomy covering $\tilde{L}$ of $L$.
Fixing a base point $x\in L$, one identifies $\tilde{L}$ with $G_{x}=\left\{ \gamma\in G,\, s(\gamma)=x\right\} $and
defines the representation\[
(\pi_{x}(f)\xi)(\gamma)=\intop_{\gamma_{1}\circ\gamma_{2}=\gamma}f(\gamma_{1})\xi(\gamma_{2})\qquad\forall\xi\in L^{2}(G_{x}).\]
The completion of $C_{c}^{\infty}(G,\Omega^{1/2})$ with respect to
the norm \[
||f||=\sup_{x\in M}||\pi_{x}(f)||\]
makes it into a $C^{*}$algebra $C_{r}^{*}(M,F)$. Among all elements
of the $C^{*}$ algebra, there are distinguished elements, idempotent
operators or projectors having a geometric interpretation in the foliation.
For later use, we will construct them explicitly (we follow \cite{Connes94}
sec. $II.8.\beta$ closely). Let $N\subset M$ be a compact submanifold
which is everywhere transverse to the foliation (thus $\dim(N)=\mathrm{codim}(F)$).
A small tubular neighborhood $N'$ of $N$ in $M$ defines an induced
foliation $F'$ of $N'$ over $N$ with fibers $\mathbb{R}^{k},\, k=\dim F$.
The corresponding $C^{*}$algebra $C_{r}^{*}(N',F')$ is isomorphic
to $C(N)\otimes\mathcal{K}$ with $\mathcal{K}$ the $C^{*}$ algebra
of compact operators. In particular it contains an idempotent $e=e^{2}=e^{*}$,
$e=1_{N}\otimes f\in C(N)\otimes\mathcal{K}$ , where $f$ is a minimal
projection in $\mathcal{K}$. The inclusion $C_{r}^{*}(N',F')\subset C_{r}^{*}(M,F)$
induces an idempotent in $C_{r}^{*}(M,F)$. Now we consider the range
map $t$ of the smooth holonomy groupoid $G$ defining via $t^{-1}(N)\subset G$
a submanifold. Let $\xi\in C_{c}^{\infty}(t^{-1}(N),s^{*}(\Omega^{1/2}))$
be a section (with compact support) of the bundle of half-density
$s^{*}(\Omega^{1/2})$ over $t^{-1}(N)$ so that the support of $\xi$
is in the diagonal in $G$ and \[
\intop_{t(\gamma)=y}|\xi(\gamma)|^{2}=1\qquad\forall y\in N.\]
Then the equality\[
e(\gamma)=\sum_{{s(\gamma)=s(\gamma')\atop t(\gamma')\in N}}\bar{\xi}(\gamma'\circ\gamma^{-1})\xi(\gamma')\]
defines an idempotent $e\in C_{c}^{\infty}(G,\Omega^{1/2})\subset C_{r}^{*}(M,F)$.
Thus, such an idempotent is given by a closed curve in $M$ transversal
to the foliation.

\subsubsection{Some information about the factor $I\! I\! I$ case\label{sub:factor-III-case}}

In our case of codimension-1 foliations of the 3-sphere with nontrivial
Godbillon-Vey invariant we have the result of Hurder and Katok \cite{HurKat:84}.
Then the corresponding von Neumann algebra $W(S^{3},F)$ contains
a factor $I\! I\! I$ algebra. At first we will give an overview about
the factor $I\! I\! I$.

Remember a von Neumann algebra is an involutive subalgebra $M$ of
the algebra of operators on a Hilbert space $H$ that has the property
of being the commutant of its commutant: $(M')'=M$. This property
is equivalent to saying that $M$ is an involutive algebra of operators
that is closed under weak limits. A von Neumann algebra $M$ is said
to be hyperfinite if it is generated by an increasing sequence of
finite-dimensional subalgebras. Furthermore we call $M$ a factor
if its center is equal to $\mathbb{C}$. It is a deep result of Murray
and von Neumann that every factor $M$ can be decomposed into 3 types
of factors $M=M_{I}\oplus M_{II}\oplus M_{III}$. The factor $I$
case divides into the two classes $I_{n}$ and $I_{\infty}$ with
the hyperfinite factors $I_{n}=M_{n}(\mathbb{C})$ the complex square
matrices and $I_{\infty}=\mathcal{L}(H)$ the algebra of all operators
on an infinite-dimensional Hilbert space $H$. The hyperfinite $I\! I$
factors are given by $I\! I_{1}=Cliff_{\mathbb{C}}(E)$, the Clifford
algebra of an infinite-dimensional Euclidean space $E$, and $I\! I_{\infty}=I\! I_{1}\otimes I_{\infty}$.
The case $I\! I\! I$ remained mysterious for a long time. Now we
know that there are three cases parametrized by a real number $\lambda\in[0,1]$:
$I\! I\! I_{0}=R_{W}$ the Krieger factor induced by an ergodic flow
$W$, $I\! I\! I_{\lambda}=R_{\lambda}$ the Powers factor for $\lambda\in(0,1)$
and $I\! I\! I_{1}=R_{\infty}=R_{\lambda_{1}}\otimes R_{\lambda_{2}}$
the Araki-Woods factor for all $\lambda_{1},\lambda_{2}$ with $\lambda_{1}/\lambda_{2}\notin\mathbb{Q}$.
We remark that all factor  $I\! I\! I$ cases are induced by infinite
tensor products of the other factors. One example of such an infinite
tensor space is the Fock space in quantum field theory. 

But now we are interested in an explicit construction of a factor
$I\! I\! I$ von Neumann algebra of a foliation. The interesting example
of this situation is given by the Anosov foliation $F$ of the unit
sphere bundle $V=T_{1}S$ of a compact Riemann surface $S$ of genus
$g>1$ endowed with its Riemannian metric of constant curvature $-1$.
In general the manifold $V$ is the quotient $V=G/T$ of the semi-simple
Lie group $G=PSL(2,\mathbb{R})$, the isometry group of the hyperbolic
plane $\mathbb{H}^{2}$, by the discrete cocompact subgroup $T=\pi_{1}(S)$,
and the foliation $F$ of $V$ is given by the orbits of the action
by left multiplication on $V=G/T$ of the subgroup of upper triangular
matrices of the form\[
\left(\begin{array}{cc}
1 & t\\
0 & 1\end{array}\right)\quad t\in\mathbb{R}\]
The von Neumann algebra $M=W(V,F)$ of this foliation is the (unique)
hyperfinite factor of type $I\! I\! I_{1}=R_{\ \infty}$. In the appendix
\ref{sec:Non-cobordant-foliationsS3} we describe the construction
of the codimension-1 foliation on the 3-sphere $S^{3}$. The main
ingredient of this construction is the convex polygon $P$ in the
hyperbolic plane $\mathbb{H}^{2}$ having curvature $-1$. The whole
construction don't depend on the number of vertices of $P$ but on
the volume $vol(P)$ only. Thus without loss of generality, we can
choose the even number $4g$ for $g\in\mathbb{N}$ of vertices for
$P$. As model of the hyperbolic plane we choose the usual upper half-plane
model where the group $SL(2,\mathbb{R})$ (the real M\"obius transformations)
and the hyperbolic group $PSL(2,\mathbb{R})$ (the group of all orientation-preserving
isometries of $\mathbb{H}^{2}$) act via fractional linear transformations.
Then the polygon $P$ is a fundamental polygon representing a Riemann
surface $S$ of genus $g$. Via the procedure above, we can construct
a foliation on $T_{1}S=PSL(2,\mathbb{R})/T$ with $T=\pi_{1}(S)$.
This foliation is also induced from the foliation of $T_{1}\mathbb{H}$(as
well as the foliation of the $S^{3}$) via the left action above.
The difference between the foliation on $T_{1}S$ and on $S^{3}$
is given by the different usage of the polygon $P$. Thus the von
Neumann algebra $W(S^{3},F)$ of the codimension-1 foliation of the
3-sphere contains a factor  $I\! I\! I$ algebra in agreement with
the results in \cite{HurKat:84}. Currently we don't know whether
it is the hyperfinite $I\! I\! I_{1}$ factor $R_{\infty}$. The Reeb
components of this foliation of $S^{3}$ (see appendix \ref{sec:Non-cobordant-foliationsS3})
are represented by a factor $I_{\infty}$ algebra and thus don't contribute
to the Godbillon-Vey class. Putting all things together we will get 

\begin{theorem} A small exotic $\mathbb{R}^{4}$ as represented by
a codimension-1 foliation of the 3-sphere $S^{3}$ with non-trivial
Godbillon-Vey invariant is also associated to a von Neumann algebra
$W(S^{3},F)$ induced by the foliation which contains a factor  $I\! I\! I$
algebra. \end{theorem} 

We conjecture that this factor  $I\! I\! I$ algebra is the hyperfinite
$I\! I\! I_{1}$ factor $R_{\infty}$. Now one may ask, what is the
physical meaning of the factor $I\! I\! I$? Because of the Tomita-Takesaki-theory,
factor $I\! I\! I$ algebras are deeply connected to the characterization
of equilibrium temperature states of quantum states in statistical
mechanics and field theory also known as Kubo-Martin-Schwinger (KMS)
condition. Furthermore in the quantum field theory with local observables
(see Borchers \cite{Borchers2000} for an overview) one obtains close
connections to Tomita-Takesaki-theory. For instance one was able to
show that on the vacuum Hilbert space with one vacuum vector the algebra
of local observables is a factor $I\! I\! I_{1}$ algebra. As shown
by Thiemann et. al. \cite{Thiemann2006} on a class of diffeomorphism
invariant theories there exists an unique vacuum vector. Thus the
observables algebra must be of this type.

\subsection{Cyclic cohomology and K-theoretic invariants of leaf spaces\label{sub:Cyclic-cohomology-and}}

As Connes \cite{Connes1984} showed, interesting K-theoretic invariants
of foliations can be described by Kasparov's KK theory. For completeness
we will give a short introduction. Then we will apply the theory to
our case.

Lets begin with KK theory. Let $A$ and $B$ be graded $C^{*}$-algebras.
$\mathbb{E}(A,B)$ is the set of all triples $(E,\phi,F)$, where
$E$ is a countably generated graded Hilbert module over $B$, $\phi$
is a graded $*$-homomorphism from $A$ to the bounded operators $\mathbb{B}(E)$
over $E$, and $F$ is an operator in $\mathbb{B}(E)$ of degree $1$,
such that $[F,\phi(a)]$, $(F^{2}-1)\phi(a)$, and $(F-F^{*})\phi(a)$
are all compact operators (i.e. in $\mathbb{K}(E)$) for all $a\in A$
(i.e. $F$ is a Fredholm operator). The elements of $\mathbb{E}(A,B)$
are called Kasparov modules for $(A,B)$. $\mathbb{D}(A,B)$ is the
set of triples in$\mathbb{E}(A,B)$ for which$[F,\phi(a)]$, $(F^{2}-1)\phi(a)$,
and $(F-F^{*})\phi(a)$ are $0$ for all $a$. The elements of $\mathbb{D}(A,B)$
are called degenerate Kasparov modules. Then one defines a homotopy
$\sim_{h}$ between two triples $(E_{i},\phi_{i},F_{i})$ for $i=0,1$
by a triple in $\mathbb{E}(A,[0,1]\times B)$ relating the two triples
in an obvious way. The KK groups are given by \[
KK(A,B)=\mathbb{E}(A,B)/\sim_{h}\]
the equivalence classes. Especially we get for $A=\mathbb{C}$ the
usual K homology $KK(\mathbb{C},B)=K_{0}(B)$ and for $B=\mathbb{C}$
the K theory $KK(A,\mathbb{C})=K^{0}(A)=K(A)$. A delicate part of
KK theory is the existence of a intersection (or cup) product $\boxtimes$
so that elements in $KK(A,C)$ and $KK(C,B)$ combine to an element
in $KK(A,B)$. See chapter VIII in the nice book \cite{Bla:86}.

Let $\mathcal{E}$ be the $C^{*}$module on $C^{*}(V,F)$ of a foliation
$F$ for a manifold $V$ (associated to the von Neumann algebra $W(V,F)$).
Let $N\stackrel{i}{\to V}$ be a transversal of the foliation constructed
above (i.e. a submanifold with $\dim N=codim(F)$ transversal to the
foliation). It defines a tranverse bundle which is associated to a
bundle $\tau$ over the classifying space $BG$ of the holonomy groupoid
$G$ of the foliation $(V,F)$. Using KK theory we can define the
K theory $K(V/F)$ of the leaf space $V/F$ via the $C^{*}$algebra
$C^{*}(V,F)$. By geometric methods Baum and Connes were able to construct
a purely geometric group $K_{*,\tau}(BG)$ together with an isomorphism
$\mu:K_{*,\tau}(BG)\to K^{*}(V/F)$ (analytic assembly map, see \cite{Connes1984}).
On $V$ the foliation determines a Haefliger structure $\Gamma$ via
the holonomy, i.e. we have a continuous map (unique up to homotopy)
$BG\to B\Gamma$ mapping elements of $K_{*,\tau}(BG)$ to elements
in $K_{*,\tau}(B\Gamma)$. The last K group is related to the usual
cohomology via Chern-Weil theory. 

Now we specialize to the codimension-1 foliation over $S^{3}$. Then
we have especially the relation between $K_{3,\tau}(B\Gamma_{1})$
and $H^{3}(B\Gamma_{1},\mathbb{R})$ and by the analytic assembly
map $\mu$ above, we have\[
K_{3,\tau}(B\Gamma_{1})\to K_{3,\tau}(BG)\stackrel{\mu}{\to}K^{3}(S^{3}/F)=K^{1}(S^{3}/F)\]
where the last equality is the Bott isomorphism. Via Connes pairing
between K theory and cyclic cohomology we have a map\[
K_{3,\tau}(B\Gamma_{1})\to HC^{1}(C^{*}(S^{3},F))\]
to the cyclic cohomology of the foliation. Thus we have shown 

\begin{theorem}

Let a small exotic $\mathbb{R}^{4}$ be determined by a codimension-1
foliation on the 3-sphere $S^{3}$. Every such foliation defines an
element in the K theory $K_{3,\tau}(B\Gamma_{1})$ determining an
element in the cyclic cohomology $HC^{1}(C^{*}(S^{3},F))$ of the
foliation (the transverse fundamental class).

\end{theorem}

It is interesting to note that we already had found an interpretation
of an element in the cyclic cohomology. In \cite{AssRos:05} we studied
the change of the connection by changing the smoothness structure
on a compact manifold to get a singular connection representing the
change. This singular connection had a close connection to a cyclic
cohomology class used to proof that the set of singular forms builds
an algebra, the Temperley-Lieb algebra or the factor $I\! I_{1}$
algebra. In subsection \ref{sub:Temperley-Lieb-algebra-foliation}
 we will show that there is a subfactor in the von Neumann algebra
$W(S^{3},F)$ of the foliation which contains a factor $I\! I_{1}$
algebra.

\section{The connection between exotic smoothness and quantization\label{sec:Quantization}}

In this section we describe a deep relation between quantization and
the codimension-1 foliation of the $S^{3}$ determining the smoothness
structure on a small exotic $\mathbb{R}^{4}$.

\subsection{From exotic smoothness to operator algebras\label{sub:From-exotic-smoothness}}

In subsection \ref{sub:smooth-holonomy-groupoid} we constructed (following
Connes \cite{Connes94}) the smooth holonomy groupoid of a foliation
$F$ and its operator algebra $C_{r}^{*}(M,F)$. The correspondence
between a foliation and the operator algebra (as well as the von Neumann
algebra) is visualized by table \ref{tab:relation-foliation-operator}.
\begin{table}
\begin{tabular}{|c|c|}
\hline 
Foliation & Operator algebra\tabularnewline
\hline
\hline 
leaf & operator\tabularnewline
\hline 
closed curve transversal to foliation & projector (idempotent operator)\tabularnewline
\hline 
holonomy & linear functional (state)\tabularnewline
\hline 
local chart & center of algebra\tabularnewline
\hline
\end{tabular}

\caption{relation between foliation and operator algebra\label{tab:relation-foliation-operator}}

\end{table}
 As extract of our previous paper \cite{AsselmeyerKrol2009}, we obtained
a relation between exotic $\mathbb{R}^{4}$'s and codimension-1 foliations
of the 3-sphere $S^{3}$. Furthermore we showed in subsection \ref{sub:factor-III-case}
that the codimension-1 foliation of $S^{3}$ consists partly of Anosov-like
foliations and Reeb foliations. Using Tomita-Takesaki-theory, one
has a continuous decomposition (as crossed product) of any factor
$I\! I\! I$ algebra $M$ into a factor $I\! I_{\infty}$ algebra
$N$ together with a one-parameter group%
\footnote{The group $\mathbb{R}_{+}^{*}$ is the group of positive real numbers
with multiplication as group operation also known as Pontrjagin dual.%
} $\left(\theta_{\lambda}\right)_{\lambda\in\mathbb{R}_{+}^{*}}$ of
automorphisms $\theta_{\lambda}\in Aut(N)$ of $N$, i.e. one obtains

\[
M=N\rtimes_{\theta}\mathbb{R}_{+}^{*}\quad.\]
But that means, there is a foliation induced from the foliation of
the $S^{3}$ producing this $I\! I_{\infty}$ factor. As we saw in
subsection \ref{sub:factor-III-case} one has a codimension-1 foliation
$F$ as part of the foliation of the $S^{3}$ whose von Neumann algebra
is the hyperfinite factor $I\! I\! I_{1}$. Connes \cite{Connes94}
(in section I.4 page 57ff) constructed the foliation $F'$ canonically
associated to $F$ having the factor $I\! I_{\infty}$ as von Neumann
algebra. In our case it is the horocycle flow: Let $P$ the polygon
on the hyperbolic space $\mathbb{H}^{2}$ determining the foliation
of the $S^{3}$ (see appendix \ref{sec:Non-cobordant-foliationsS3}).
$P$ is equipped with the hyperbolic metric $2|dz|/(1-|z|^{2})$ together
with the collection $T_{1}P$ of unit tangent vectors to $P$. A horocycle
in $P$ is a circle contained in $P$ which touches $\partial P$
at one point. Then the horocycle flow $T_{1}P\to T_{1}P$ is the flow
moving an unit tangent vector along a horocycle (in positive direction
at unit speed). As above the polygon $P$ determines a surface $S$
of genus $g>1$ with abelian torsion-less fundamental group $\pi_{1}(S)$
so that the homomorphism $\pi_{1}(S)\to\mathbb{R}$ determines an
unique (ergodic invariant) Radon measure. Finally the horocycle flow
determines a factor $I\! I_{\infty}$ foliation associated to the
factor $I\! I\! I_{1}$ foliation. We remark for later usage that
this foliation is determined by a set of closed curves (the horocycles). 

Using results of our previous papers, we have the following picture:

\begin{enumerate}
\item Every small exotic $\mathbb{R}^{4}$ is determined by a codimension-1
foliation (unique up to cobordisms) of some homology 3-sphere $\Sigma$
(as boundary $\partial A=\Sigma$ of a contractable submanifold $A\subset\mathbb{R}^{4}$,
the Akbulut cork).
\item This codimension-1 foliation on $\Sigma$ determines via surgery along
a link uniquely a codimension-1 foliation on the 3-sphere and vice
verse.
\item This codimension-1 foliation $(S^{3},F)$ on $S^{3}$ has a leaf space
which is determined by the von Neumann algebra $W(S^{3},F)$ associated
to the foliation.
\item The von Neumann algebra $W(S^{3},F)$ contains a hyperfinite factor
$I\! I\! I_{1}$ algebra as well as a factor $I_{\infty}$ algebra
coming from the Reeb foliations.
\end{enumerate}
Thus by this procedure we get a noncommutative algebra from an exotic
$\mathbb{R}^{4}$. The relation to the quantum theory will be discussed
now. We remark that we have already a quantum theory represented by
the von Neumann algebra $W(S^{3},F)$. Thus we are in the strange
situation to construct a (classical) Poisson algebra together with
a quantization to get an algebra which we already have.

\subsection{The observable algebra and Poisson structure\label{sub:The-observable-algebra}}

In this section we will describe the formal structure of a classical
theory coming from the algebra of observables using the concept of
a Poisson algebra. In quantum theory, an observable is represented
by a hermitean operator having the spectral decomposition via projectors
or idempotent operators. The coefficient of the projector is the eigenvalue
of the observable or one possible result of a measurement. At least
one of these projectors represent (via the GNS representation) a quasi-classical
state. Thus to construct the substitute of a classical observable
algebra with Poisson algebra structure we have to concentrate on the
idempotents in the $C^{*}$ algebra.

In subsection \ref{sub:smooth-holonomy-groupoid}, an idempotent was
constructed in the $C^{*}$ algebra of the foliation and geometrically
interpreted as closed curve transversal to the foliation. Such
a curve meets every leaf in a finite number of points. Furthermore
the 3-sphere $S^{3}$ is embedded in the some 4-space with the tubular
neighborhood $S^{3}\times[0,1]$. Then we have a thickened curve $S^{1}\times[0,1]$
or a closed curve on a surface. Thus we have to consider closed curves
in surfaces. Now we will see that the
set of closed curves on a surface has the structure of a Poisson algebra. 

Let us start with the definition of a Poisson algebra. Let $P$ be
a commutative algebra with unit over $\mathbb{R}$ or $\mathbb{C}$.
A \emph{Poisson bracket} on $P$ is a bilinearform $\left\{ \:,\:\right\} :P\otimes P\to P$
fulfilling the following 3 conditions:

\begin{itemize}
\item anti-symmetry $\left\{ a,b\right\} =-\left\{ b,a\right\} $
\item Jacobi identity $\left\{ a,\left\{ b,c\right\} \right\} +\left\{ c,\left\{ a,b\right\} \right\} +\left\{ b,\left\{ c,a\right\} \right\} =0$
\item derivation $\left\{ ab,c\right\} =a\left\{ b,c\right\} +b\left\{ a,c\right\} $.
\end{itemize}
Then a \emph{Poisson algebra} is the algebra $(P,\{\,,\,\})$. Now
we consider a surface $S$ together with a closed curve $\gamma$.
Additionally we have a Lie group $G$ given by the isometry group.
The closed curve is one element of the fundamental group $\pi_{1}(S)$.
From the theory of surfaces we know that $\pi_{1}(S)$ is a free abelian
group. Denote by $Z$ the free $\mathbb{K}$-module ($\mathbb{K}$
a ring with unit) with the basis $\pi_{1}(S)$, i.e. $Z$ is a freely
generated $\mathbb{K}$-modul. Recall Goldman's definition of the
Lie bracket in $Z$ (see \cite{Goldman1984}). For a loop $\gamma:S^{1}\to S$
we denote its class in $\pi_{1}(S)$ by $\left\langle \gamma\right\rangle $.
Let $\alpha,\beta$ be two loops on $S$ lying in general position.
Denote the (finite) set $\alpha(S^{1})\cap\beta(S^{1})$ by $\alpha\#\beta$.
For $q\in\alpha\#\beta$ denote by $\epsilon(q;\alpha,\beta)=\pm1$
the intersection index of $\alpha$ and $\beta$ in $q$. Denote by
$\alpha_{q}\beta_{q}$ the product of the loops $\alpha,\beta$ based
in $q$. Up to homotopy the loop $(\alpha_{q}\beta_{q})(S^{1})$ is
obtained from $\alpha(S^{1})\cup\beta(S^{1})$ by the orientation
preserving smoothing of the crossing in the point $q$. Set \begin{equation}
[\left\langle \alpha\right\rangle ,\left\langle \beta\right\rangle ]=\sum_{q\in\alpha\#\beta}\epsilon(q;\alpha,\beta)(\alpha_{q}\beta_{q})\quad.\label{eq:Lie-bracket-loops}\end{equation}
According to Goldman \cite{Goldman1984}, Theorem 5.2, the bilinear
pairing $[\,,\,]:Z\times Z\to Z$ given by (\ref{eq:Lie-bracket-loops})
on the generators is well defined and makes $Z$ to a Lie algebra.
The algebra $Sym(Z)$ of symmetric tensors is then a Poisson algebra
(see Turaev \cite{Turaev1991}).

The whole approach seems natural for the construction of the Lie algebra
$Z$ but the introduction of the Poisson structure is an artificial
act. From the physical point of view, the Poisson structure is not
the essential part of classical mechanics. More important is the algebra
of observables, i.e. functions over the configuration space forming
the Poisson algebra. Thus we will look for the algebra of observables
in our case. For that purpose, we will look at geometries over the
surface. By the uniformization theorem of surfaces, there is three
types of geometrical models: spherical $S^{2}$, Euclidean $\mathbb{E}^{2}$
and hyperbolic $\mathbb{H}^{2}$. Let $\mathcal{M}$ be one of these
models having the isometry group $Isom(\mathcal{M})$. Consider a
subgroup $H\subset Isom(\mathcal{M})$ of the isometry group acting
freely on the model $\mathcal{M}$ forming the factor space $\mathcal{M}/H$.
Then one obtains the usual (closed) surfaces $S^{2}$, $\mathbb{R}P^{2}$,
$T^{2}$ and its connected sums like the surface of genus $g$ ($g>1$).
For the following construction we need a group $G$ containing the
isometry groups of the three models. Furthermore the surface $S$
is part of a 3-manifold and for later use we have to demand that $G$
has to be also a isometry group of 3-manifolds. According to Thurston
\cite{Thu:97} there are 8 geometric models in dimension 3 and the
largest isometry group is the hyperbolic group $PSL(2,\mathbb{C})$
isomorphic to the Lorentz group $SO(3,1).$ It is known that every
representation of $PSL(2,\mathbb{C})$ can be lifted to the spin group
$SL(2,\mathbb{C})$. Thus the group $G$ fulfilling all conditions
is identified with $SL(2,\mathbb{C})$. This choice fits very well
with the 4-dimensional picture.

Now we introduce a principal $G$ bundle on $S$, representing a geometry
on the surface. This bundle is induced from a $G$ bundle over $S\times[0,1]$
having always a flat connection. Alternatively one can consider a
homomorphism $\pi_{1}(S)\to G$ represented as holonomy functional\[
hol(\omega,\gamma)=\mathcal{P}\exp\left(\int\limits _{\gamma}\omega\right)\in G\]
with the path ordering operator $\mathcal{P}$ and $\omega$ as flat
connection (i.e. inducing a flat curvature $\Omega=d\omega+\omega\wedge\omega=0$).
This functional is unique up to conjugation induced by a gauge transformation
of the connection. Thus we have to consider the conjugation classes
of maps\[
hol:\pi_{1}(S)\to G\]
forming the space $X(S,G)$ of gauge-invariant flat connections of
principal $G$ bundles over $S$. Now (see \cite{Skovborg2006}) we
can start with the construction of the Poisson structure on $X(S,G).$
The construction based on the Cartan form as the unique bilinearform
of a Lie algebra. As discussed above we will use the Lie group $G=SL(2,\mathbb{C})$
but the whole procedure works for every other group too. Now we consider
the standard basis\[
X=\left(\begin{array}{cc}
0 & 1\\
0 & 0\end{array}\right)\quad,\qquad H=\left(\begin{array}{cc}
1 & 0\\
0 & -1\end{array}\right)\quad,\qquad Y=\left(\begin{array}{cc}
0 & 0\\
1 & 0\end{array}\right)\]
of the Lie algebra $sl(2,\mathbb{C})$ with $[X,Y]=H,\,[H,X]=2X,\,[H,Y]=-2Y$.
Furthermore there is the bilinearform $B:sl_{2}\otimes sl_{2}\to\mathbb{C}$
written in the standard basis as \[
\left(\begin{array}{ccc}
0 & 0 & -1\\
0 & -2 & 0\\
-1 & 0 & 0\end{array}\right)\]
Now we consider the holomorphic function $f:SL(2,\mathbb{C})\to\mathbb{C}$
and define the gradient $\delta_{f}(A)$ along $f$ at the point $A$
as $\delta_{f}(A)=Z$ with $B(Z,W)=df_{A}(W)$ and \[
df_{A}(W)=\left.\frac{d}{dt}f(A\cdot\exp(tW))\right|_{t=0}\quad.\]
The calculation of the gradient $\delta_{tr}$ for the trace $tr$
along a matrix \[
A=\left(\begin{array}{cc}
a_{11} & a_{12}\\
a_{21} & a_{22}\end{array}\right)\]
 is given by\[
\delta_{tr}(A)=-a_{21}Y-a_{12}X-\frac{1}{2}(a_{11}-a_{22})H\quad.\]
Given a representation $\rho\in X(S,SL(2,\mathbb{C}))$ of the fundamental
group and an invariant function $f:SL(2,\mathbb{C})\to\mathbb{R}$
extendable to $X(S,SL(2,\mathbb{C}))$. Then we consider two conjugacy
classes $\gamma,\eta\in\pi_{1}(S)$ represented by two transversal
intersecting loops $P,Q$ and define the function $f_{\gamma}:X(S,SL(2,\mathbb{C})\to\mathbb{C}$
by $f_{\gamma}(\rho)=f(\rho(\gamma))$. Let $x\in P\cap Q$ be the
intersection point of the loops $P,Q$ and $c_{x}$ a path between
the point $x$ and the fixed base point in $\pi_{1}(S)$. The we define
$\gamma_{x}=c_{x}\gamma c_{x}^{-1}$ and $\eta_{x}=c_{x}\eta c_{x}^{-1}$.
Finally we get the Poisson bracket\[
\left\{ f_{\gamma},f'_{\eta}\right\} =\sum_{x\in P\cap Q}sign(x)\: B(\delta_{f}(\rho(\gamma_{x})),\delta_{f'}(\rho(\eta_{x})))\quad,\]
where $sign(x)$ is the sign of the intersection point $x$. Thus
the space $X(S,SL(2,\mathbb{C}))$ has a natural Poisson structure
(induced by the bilinear form on the group) and the Poisson algebra
\emph{$(X(S,SL(2,\mathbb{C}),\left\{ \,,\,\right\} )$} of complex
functions over them is the algebra of observables.

\subsection{Drinfeld-Turaev Quantization\label{sub:Drinfeld-Turaev-Quantization}}

Now we introduce the ring $\mathbb{C}[[h]]$ of formal polynomials
in $h$ with values in $\mathbb{C}$. This ring has a topological
structure, i.e. for a given power series $a\in\mathbb{C}[[h]]$ the
set $a+h^{n}\mathbb{C}[[h]]$ forms a neighborhood. Now we define
a \emph{Quantization} of a Poisson algebra $P$ as a $\mathbb{C}[[h]]$
algebra $P_{h}$ together with the $\mathbb{C}$-algebra isomorphism
$\Theta:P_{h}/hP\to P$ so that

\begin{itemize}
\item the modul $P_{h}$ is isomorphic to $V[[h]]$ for a $\mathbb{C}$
vector space $V$
\item let $a,b\in P$ and $a',b'\in P_{h}$ be $\Theta(a)=a'$, $\Theta(b)=b'$
then\[
\Theta\left(\frac{a'b'-b'a'}{h}\right)=\left\{ a,b\right\} \]

\end{itemize}
One speaks of a deformation of the Poisson algebra by using a deformation
parameter $h$ to get a relation between the Poisson bracket and the
commutator.

Now we have the problem to find the deformation of the Poisson algebra
$(X(S,SL(2,\mathbb{C})),\left\{ \,,\,\right\} )$. The solution to
this problem can be found via two steps: at first find another description
of the Poisson algebra by a structure with one parameter at a special
value and secondly vary this parameter to get the deformation. Fortunately
both problems were already solved (see \cite{Turaev1989,Turaev1991}).
The solution of the first problem is expressed in the theorem: \emph{The
Skein modul $K_{-1}(S\times[0,1])$ (i.e. $t=-1$) has the structure
of an algebra isomorphic to the Poisson algebra $(X(S,SL(2,\mathbb{C}),\left\{ \,,\,\right\} )$.}
(see also \cite{BulPrzy:1999,Bullock1999}) Then we have also the
solution of the second problem: \emph{The skein algebra $K_{t}(S\times[0,1])$
is the quantization of the Poisson algebra $(X(S,SL(2,\mathbb{C}),\left\{ \,,\,\right\} )$
with the deformation parameter $t=\exp(h/4)$.}(see also \cite{BulPrzy:1999})\emph{
}To understand these solutions we have to introduce the skein module
$K_{t}(M)$ of a 3-manifold $M$ (see \cite{PrasSoss:97}).

\begin{figure}
\centering
\resizebox{0.5\textwidth}{!}{
\includegraphics{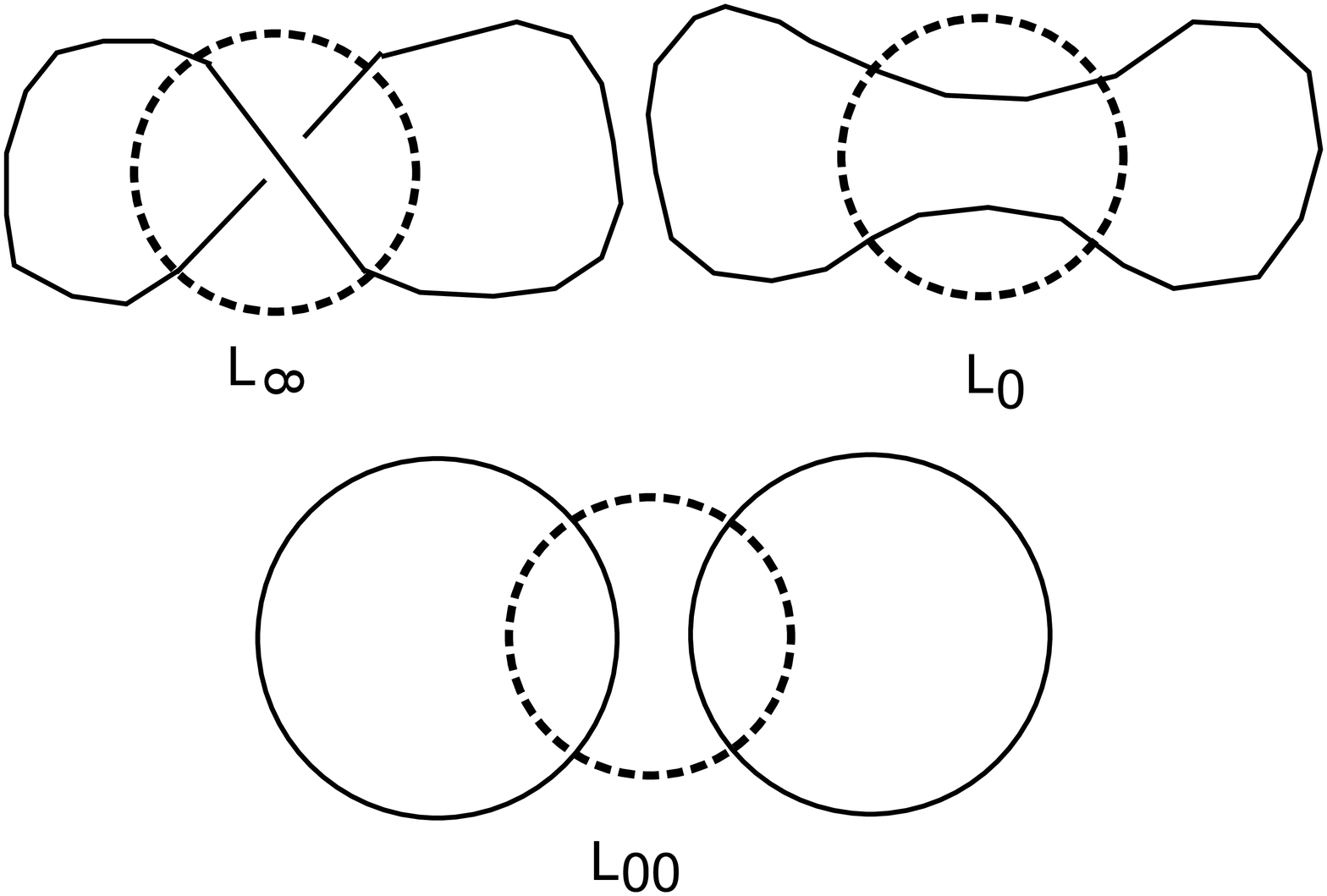}
}

\caption{crossings $L_{\infty},L_{o},L_{oo}$}
\label{fig:skein-crossings}

\end{figure}

For that purpose we consider the set of links $\mathcal{L}(M)$ in
$M$ up to isotopy and construct the vector space $\mathbb{C}\mathcal{L}(M)$
with basis $\mathcal{L}(M)$. Then one can define $\mathbb{C}\mathcal{L}[[t]]$
as ring of formal polynomials having coefficients in $\mathbb{C}\mathcal{L}(M)$.
Now we consider the link diagram of a link, i.e. the projection of
the link to the $\mathbb{R}^{2}$ having the crossings in mind. Choosing
a disk in $\mathbb{R}^{2}$ so that one crossing is inside this disk.
Differ three links by the three crossings $L_{oo},L_{o},L_{\infty}$
(see figure \ref{fig:skein-crossings}) inside of the disk then these
links are skein related. %
Then in $\mathbb{C}\mathcal{L}[[t]]$ one writes the skein relation%
\footnote{The relation depends on the group $SL(2,\mathbb{C})$.%
} $L_{\infty}-tL_{o}-t^{-1}L_{oo}$. Furthermore let $L\sqcup O$ be
the disjoint union of the link with a circle then one writes the framing
relation $L\sqcup O+(t^{2}+t^{-2})L$. Let $S(M)$ be the smallest
submodul of $\mathbb{C}\mathcal{L}[[t]]$ containing both relations,
then we define the Kauffman bracket skein modul by $K_{t}(M)=\mathbb{C}\mathcal{L}[[t]]/S(M)$.
We list the following general results about this modul:

\begin{itemize}
\item The modul $K_{-1}(M)$ for $t=-1$ is a commutative algebra.
\item Let $S$ be a surface then $K_{t}(S\times[0,1])$ caries the structure
of an algebra.
\end{itemize}
The algebra structure of $K_{t}(S\times[0,1])$ can be simple seen
by using the diffeomorphism between the sum $S\times[0,1]\cup_{S}S\times[0,1]$
along $S$ and $S\times[0,1]$. Then the product $ab$ of two elements
$a,b\in K_{t}(S\times[0,1])$ is a link in $S\times[0,1]\cup_{S}S\times[0,1]$
corresponding to a link in $S\times[0,1]$ via the diffeomorphism.
The algebra $K_{t}(S\times[0,1])$ is in general non-commutative for
$t\not=-1$. 

For the following we will omit the interval $[0,1]$ and denote the
skein algebra by $K_{t}(S)$. Then we have:

\begin{theorem} The skein algebra $(K_{t}(S),[\,,\,])$ is the quantization
of the Poisson algebra $(X(S,SL(2,\mathbb{C}),\left\{ \,,\,\right\} )$
with deformation parameter $t=\exp(h/4)$. \end{theorem}

\emph{Ad hoc} the skein algebra is not directly related to the foliation.
We used only the fact that there is an idempotent in the $C^{*}$
algebra represented by a closed curve. It is more satisfying to obtain
a direct relation between both construction. Then the von Neumann
algebra of the foliation is the result of a quantization in the physical
sense. This construction is left for the next subsection.

\subsection{Temperley-Lieb algebra and the operator algebra of the foliation\label{sub:Temperley-Lieb-algebra-foliation}}

In this subsection we will describe a direct relation between the
skein algebra and the factor $I\! I\! I_{1}$ constructed above. At
first we will summarize some of the results above. 

\begin{enumerate}
\item The foliation of the 3-sphere $S^{3}$ has non-trivial Godbillon-Vey
class. The corresponding von Neumann algebra must contain a factor
$I\! I\! I$ algebra.
\item We conjectured that the von Neumann algebra is the hyperfinite factor
$I\! I\! I_{1}$ determined by a factor $I\! I_{\infty}$ algebra
via Tomita-Takesaki theory.
\item In the von Neumann algebra there are idempotent operators given by
closed curves in the foliation.
\item The set of closed curves carries the structure of the Poisson algebra
whose quantization is the skein algebra determined by knots and links.
Thus the skein algebra can be seen as a quantization of the fundamental
group.
\end{enumerate}
Thus our main goal in this subsection should be a direct relation
between a suitable skein algebra and the von Neumann algebra of the
foliation. As a first step we remark that a factor $I\! I_{\infty}$
algebra is the tensor product $I\! I_{\infty}=I\! I_{1}\otimes I_{\infty}$.
Thus the main factor is given by the $I\! I_{1}$ factor, i.e. a von
Neumann algebra with finite trace. From the point of view of invariants,
both factors $I\! I_{\infty}$ and $I\! I_{1}$ are Morita-equivalent
leading to the same K-theoretic invariants. 

Now we are faced with the question: Is there any skein algebra isomorphic
to the factor $I\! I_{1}$ algebra? Usually the skein algebra is finite
or finitely generated (as module over the first homology group). Thus
we have to construct a finite algebra reconstructing the factor $I\! I_{1}$
in the limit. Following the theory of Jones \cite{Jon:83}, one uses
a tower of Temperley-Lieb algebras as generated by projection (or
idempotent) operators. Thus, if we are able to show that a skein algebra
constructed from the foliation is isomorphic to the Temperley-Lieb
algebra then we have constructed the factor $I\! I_{1}$ algebra. 

\begin{figure}
\centering
\resizebox{0.5\textwidth}{!}{
\includegraphics{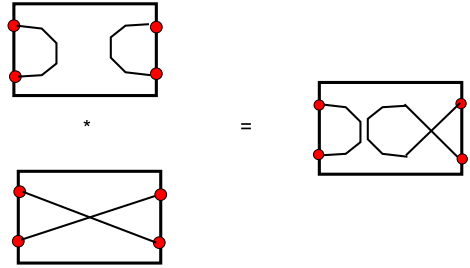}
}
\caption{product structure} 
\label{fig:product-structure}

\end{figure}

\begin{figure}
\centering
\resizebox{0.5\textwidth}{!}{
\includegraphics{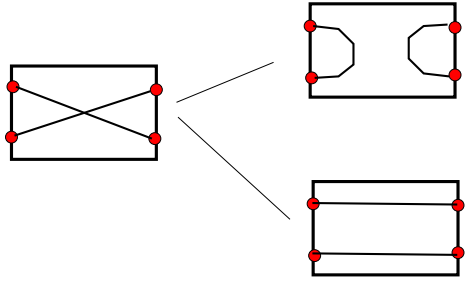}
}
\caption{resolution of the flow singularities}
\label{fig:resolution-sing}

\end{figure}

For the construction we go back to factor $I\! I_{\infty}$ foliation
discussed above and identified as the horocycle foliation. Let $P$
the polygon with hyperbolic metric used in subsection \ref{sub:factor-III-case}
and in appendix \ref{sec:Non-cobordant-foliationsS3}. Remember a
horocycle is a circle in the interior of $P$ touching the boundary
at one point. Now we consider the flow in $T_{1}P$ along a horocycle
with unit speed which induces a codimension-1 foliation in $T_{1}P$.
Before we will describe the details of the approach, we will present
the main idea. The horocycle foliation is parametrized by the set
of horocycles on $P$. Every horocycle meets the boundary of $P$
at one point, which we mark. %
\begin{figure}
\centering
\resizebox{0.5\textwidth}{!}{
\includegraphics{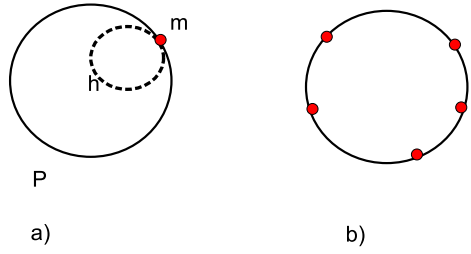}
}
\caption{fig. a) an example for a horocycle $h$ on $P$ and fig. b) marked
points on the boundary of the polygon $P$}
\label{fig:horocycle}
\end{figure}
Using the horocycles we obtain a flow for every pair of marked points.
The set of unit tangent vectors labels the leaves of the foliation
and can be described by curves between the marked points. Then we
group the marked points and assume that we have the same number of
marked points on the left and on the right side of $P$. It is obvious
that two polygons with the same number of marked points on one side
can be put together (product operation, see figure \ref{fig:product-structure}).
 Special attention should be given to flows between two marked points
of the same side. Such flows appear if one resolves the singularities
of the flow (see figure \ref{fig:resolution-sing}). Then we obtain
the relations of the Temperley-Lieb algebra which is the skein algebra
of a disk with marked points on the boundary (see \cite{PrasSoss:97}
§ 26.9.). As Jones \cite{Jon:83} showed: the limit case is the factor
$I\! I_{1}$. 

\begin{theorem}

The leaf space of the horocycle foliation of a surface $S$ (of genus
$g>1$) represented by a horocycle foliation of a polygon $P$ in
$\mathbb{H}^{2}$ can be given the structure of an algebra by connected
sum of the polygons. This algebra is isomorphic to the hyperfinite
factor $I\! I_{1}$ algebra given as tower of Temperley-Lieb algebras
(and equal to the skein algebra of a marked disk).

\end{theorem}

After the general overview we will now present the details. Given
a polygon $P$ as covering space of a surface $S$ (of genus $g>1$)
with nonpositive curvature. Denote by $\gamma_{v}$ the geodesic with
initial tangent vector $v$ and by $dist(\gamma_{v}(t),\gamma_{w}(t))$
the distance between two points on two curves. We call the two tangent
vectors $v,w$ of the cover $P$ asymptotic if the distance $dist(\gamma_{v}(t),\gamma_{w}(t))$
is bounded as $t\to\infty$. For a unit tangent vector $v\in T_{1}P$
define the Busemann function $b_{v}:P\to\mathbb{R}$ by\[
b_{v}(q)=\lim_{t\to\infty}\left(dist(\gamma_{v}(t),q)-t\right)\]
This function is differentiable and the gradient $-\nabla_{q}b_{v}$
is the unique vector at $q$ asymptotic to $v$. We define alternatively
the horocycle $h(v)$ (determined by $v$) as the level set $b_{v}^{-1}(0)$.
Clearly $h(v)$ is the limit as $R\to\infty$ of the geodesic circles
of radius $R$ centered at $\gamma_{v}(R)$. Let $W(v)$ be the set
of vectors $w$ asymptotic to $v$ with footpoints on $h(v)$, i.e.\[
W(v)=\left\{ -\nabla_{q}b_{v}\,|\, q\in h(v)\right\} \:.\]
The curves $W(v),\: v\in T_{1}P$ are the leaves of the horocycle
foliation $W$ of $T_{1}P$ which can be lifted to a horocycle foliation
$W$ on $T_{1}S$. Thus the set of unit tangent vectors labels the
leaves of the foliation or the leaf space is parametrized by unit
tangent vectors. Furthermore we remark that every horocycle is also
determined by a unit tangent vector. By definition, the set of unit
tangent vectors is completely determined by curves in $P$. Therefore
consider a finite set of horocycles and mark the boundary points of
these cycles, say $m_{1},\ldots,m_{n}$. %
\begin{figure}
\centering
\resizebox{0.5\textwidth}{!}{
\includegraphics{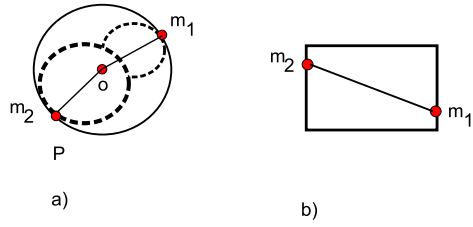}
}

\caption{a) flow line between two marked points defined via two horocycles,
b) simple picture as substitute}
\label{fig:flow-line}

\end{figure}
Then by uniqueness of the flow, there is a curve from the boundary
point $m_{1}$ in the interior of $P$ meeting a point $o$ followed
by a curve from this point $o$ to another boundary point $m_{2}$
(see figure \ref{fig:flow-line}). Thus in general we obtain curves
in $P$ going from one marked boundary point $m_{k}$ to another marked
boundary point $m_{l}$. 

Now we will construct a set of generators for these curves. Formally
we can group the marked points into two classes (see figure \ref{fig:product-structure}),
lying on the left or right (we assume w.l.o.g. an even number). We
remark that one can introduce a product like in figure \ref{fig:product-structure}.
If we connect all marked points with each other then we obtain flows
intersecting each other, i.e. we obtain singular flows. But the singularities
or intersection points can be solved to get non-singular flows.%
 The figure \ref{fig:resolution-sing} shows the method%
\footnote{The method was used in the theory of finite knot invariants (Vassiliev
invariants) and is known as STU relation.%
}. %
\begin{figure}
\centering
\resizebox{0.5\textwidth}{!}{
\includegraphics{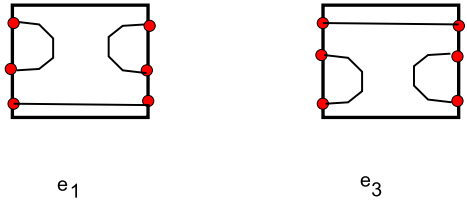}
}
\caption{example for generators $e_{1}$ and $e_{3}$ of the Temperley-Lieb
algebra}
 \label{fig:generators-TL-alg}

\end{figure}
 It is easy to see that we obtain the generators (see figure \ref{fig:generators-TL-alg})
of the skein algebra of a disk with marked points. Lets denote these
generators by $e_{i}$. By simple graphical manipulation (see \cite{PrasSoss:97})
we obtain the relations:\begin{eqnarray}
e_{i}^{2}=e_{i}\,, & e_{i}e_{j}=e_{j}e_{i}\,:\,|i-j|>1,\nonumber \\
e_{i}e_{i+1}e_{i}=e_{i}\,, & e_{i+1}e_{i}e_{i+1}=e_{i+1}\,,\, e_{i}^{*}=e_{i}\label{Jones-algebra}\end{eqnarray}
where the $*$operation is a simple change of the orientation of the
curves. Now we need a delicate argument to get the desired relation
to factor $I\! I_{1}$. Up to now we have only countable infinite
curves between marked points on the disk. To represent the whole set
of unit tangent vectors, we need all possible curves on the disk.
We can approximate between two curves represented by two generators
$e_{k},e_{n}$ using the linear combination $ae_{k}+be_{n}$ and introducing
the sum operation. The coefficients $a,b$ are complex numbers representing
a 2-dimensional deformation of the curve. The set of countable infinite
generators $e_{1},e_{2},\ldots$ together with its linear span $\mathcal{L}(e_{1},e_{2},\ldots)$
is dense in the set of all possible curves, i.e. a Cauchy sequences
of curves in $\mathcal{L}(e_{1},e_{2},\ldots)$ approximate every
arbitrary curve. Thus we have constructed the factor $I\! I_{1}$
algebra as skein algebra.

We will finish this subsection with one remark. In the factor Temperley-Lieb
algebra there is an unique idempotent operator, the Jones-Wenzl idempotent,
which is related to Connes idempotent operator in the operator algebra
of the foliation by our construction.

\subsection{Casson handle and operator algebra\label{sub:Casson-handle-and}}

For amusement we will also discuss a direct relation between the Casson
handle and the factor $I\! I_{1}$ algebra represented by the Temperley-Lieb
algebra. This will support the result of our previous subsection.
The following discussion is adapted from the book \cite{Asselmeyer2007}.

According to Freedman (\cite{Fre:82} p.393), a Casson handle is represented
by a labeled finitely-branching tree $Q$ with base point $\star$,
having all edge paths infinitely extendable away from $\star$ Each
edge should be given a label $+$ or $-$ and each vertex corresponds
to a kinky handle where the self-plumbing number of that kinky handle
equals the number of branches leaving the vertex. The sign on each
branch corresponds to the sign of the associated self plumbing. 

Now we are ready to describe the relation to the algebra of the previous
subsection. Lets take start with the tree $Q$. Every path in this
tree represents one leave in the foliation of the $S^{3}$. Two different
paths in the tree represent two different leaves in the foliation.
Then we have to consider two paths in the tree $Q$, the reference
path for the given leaf and a path for the another leaf of the foliation.
Thus, a pair of two paths corresponds to one element of the algebra.
To recover all relations of such pairs of paths in a tree we have
to consider the so-called string algebra according to Ocneanu \cite{Ocn:88}.
For that purpose we define a non-negative function $\mu:Edges\to{\mathbb{C}}$
together with the adjacency matrix $\triangle$ acting on $\mu$ by
\[
\triangle\mu(x)=\sum\limits _{{v\in Edges\atop {s(v)=x\atop r(v)=y}}}\mu(y)\]
 where $s(v)$ and $r(v)$ denote the source and the range of an edge
$v$. A path in the tree is a succession of edges $\xi=(v_{1},v_{2},\ldots,v_{n})$
where $r(v_{i})=s(v_{i+1})$ and we write $\tilde{v}$ for the edge
$v$ with the reversed orientation. Then, a string on the tree is
a pair of paths $\rho=(\rho_{+},\rho_{-})$, with $s(\rho_{+})=s(\rho_{-})$,
$r(\rho_{+})\sim r(\rho_{-})$ which means that $r(\rho_{+})$ and
$r(\rho_{-})$ ending on the same level in the tree and $\rho_{+},\rho_{-}$
have equal lengths i.e. $|\rho_{+}|=|\rho_{-}|$ expressing the previous
described property $r(\rho_{+})\sim r(\rho_{-})$ too. Now we define
an algebra $String^{(n)}$ with the linear basis of the $n$-strings,
i.e. strings with length $n$ and the additional operations: \begin{eqnarray*}
(\rho_{+},\rho_{-})\cdot(\eta_{+},\eta_{-}) & = & \delta_{\rho_{-},\eta_{+}}(\rho_{+},\eta_{-})\\
(\rho_{+},\rho_{-})^{*} & = & (\rho_{-},\rho_{+})\end{eqnarray*}
 where $\cdot$ can be seen as the concatenation of paths. We normalize
the function $\mu$ by $\mu(root)=1$. Now we choose a function $\mu$
in such a manner that \begin{equation}
\triangle\mu=\beta\mu\label{graph-laplace}\end{equation}
 for a complex number $\beta$. Then we can construct elements $e_{n}$
in the algebra $String^{(n+1)}$ by \begin{equation}
e_{n}=\sum\limits _{{|\alpha|=n-1\atop |v|=|w|=1}}\frac{\sqrt{\mu(r(v))\mu(r(w))}}{\mu(r(\alpha))}(\alpha\cdot v\cdot\tilde{v},\alpha\cdot w\cdot\tilde{w})\label{Jones-projection}\end{equation}
fulfilling the algebraic relations \begin{eqnarray}
e_{i}^{2}=e_{i}\,, & e_{i}e_{j}=e_{j}e_{i}\,:\,|i-j|>1,\nonumber \\
e_{i}e_{i+1}e_{i}=e_{i}\,, & e_{i+1}e_{i}e_{i+1}=\tau\cdot e_{i+1}\,,\, e_{i}^{*}=e_{i}\label{Jones-TL-algebra}\end{eqnarray}
where $\tau=\beta^{-2}$. The trace of the string algebra given by
\[
tr(\rho)=\delta_{\rho_{+},\rho_{-}}\beta^{-|\rho|}\mu(r(\rho))\]
 defines on $A_{\infty}=(\bigcup\limits _{n}String^{(n)},tr)$ an
inner product by $\langle x,y\rangle=tr(xy^{*})$ which is after completion
the Hilbert space $L^{2}(A_{\infty},tr)$.

Now we will determine the parameter $\tau$. Originally, Ocneanu introduce
its string algebra to classify the splittings of modules over an operator
algebra. Thus, to determine this parameter we look for the simplest
generating structure in the tree. The simplest structure in our tree
is one edge which is connected with two other edges. This graph is
represented by the following adjacence matrix \begin{eqnarray*}
\left(\begin{array}{ccc}
0 & 1 & 1\\
1 & 0 & 0\\
1 & 0 & 0\end{array}\right)\end{eqnarray*}
 having eigenvalues $0,\sqrt{2},-\sqrt{2}$. According to our definition
above, $\beta$ is given by the greatest eigenvalue of this adjacence
matrix, i.e. $\beta=\sqrt{2}$ and thus $\tau=\beta^{-2}=\frac{1}{2}$.
Then, without proof, we state that this algebra is given by the Clifford
algebra on ${\mathbb{R}}^{\infty}$, i.e. by the hyperfinite factor
$I\! I_{1}$ algebra.

\subsection{The approach via topoi\label{sub:Topoi}}

In this section we propose to consider exotic $\mathbb{R}^{4}$ allowing
for a fundamental approach to theories of physics at least in dimension
4. This is in some analogy with the approach via topos theory, recently
under active development. The topos techniques are universally valid
in any dimension, but non-classical, intuitionistic logic is an essential
part. Now we will make some additional suppositions to understand
foliation-based quantization presented above in the context of 4-dimensional
quantum field theories. 

Firstly we assume (see subsection \ref{sub:factor-III-case}): \emph{($A$)
the appearance of the factor $I\! I\! I_{1}$ in the von Neumann algebra
of the codimension-1 foliations of $S^{3}\subset\mathbb{R}^{4}$}.
Then this factor $I\! I\! I_{1}$ indicates that exotic 4-smoothness
of $\mathbb{R}^{4}$ can be connected with the regime of relativistic
local QFT or with the statistical limit of quantum systems \cite{Haag:96}. 

Let us suppose that indeed in the case of a small exotic $\mathbb{R}^{4}$:
($B$)\emph{ the factor }$I\! I\! I_{1}$\emph{, obtained from the
codimension-1 foliation of $S^{3}$ which corresponds to the exotic
$\mathbb{R}^{4}$, is the factor }$I\! I\! I_{1}$\emph{ algebra of
quantum observables of some relativistic local QFT (RQFT) defined
on standard $\mathbb{R}^{4}$ (with the Minkowskian metric}). Equivalently
one could perform a construction of a RQFT on $\mathbb{R}^{4}$ with
suitable scaling asymptotic conditions on states of the theory and
the algebra of local observables of the theory must be the factor
$I\! I\! I_{1}$. Then this factor is isomorphic to the von Neumann
algebra of the codimension-1 foliation of $S^{3}\subset\mathbb{R}^{4}$.
Note that in the case of standard smooth $\mathbb{R}^{4}$ the Godbillon-Vey
invariant vanishes, so we do not have the corresponding foliations
of $S^{3}$, hence the $I\! I\! I_{1}$ does not appear.

Moreover, we obtained the factor $I\! I\! I_{1}$, corresponding to
some nonstandard smoothing of $\mathbb{R}^{4}$, as deformation quantization
of the classical Poisson algebra underlying the observables of some
classical theory. We see that standard smoothing of $\mathbb{R}^{4}$
does not give rise to this kind of quantization, since the factor
$I\! I\! I_{1}$ algebra does not appear. So, standard smooth $\mathbb{R}^{4}$
should correspond to some classical theory. Let us suppose that: ($C$)
\emph{there exists a classical theory, defined on Euclidean or Minkowskian
$\mathbb{R}^{4}$, which determines the theory whose algebra of observables
is the Poisson algebra $(X(S,SL(2,\mathbb{C}),\left\{ \,,\,\right\} )$
of complex functions on $X(S,SL(2,\mathbb{C}))$. }

We do not discuss the details of the determination nor the specific
theories appearing in the assumptions ($B$) and ($C$). We just assume
the existence of such theories and are interested in their algebraic
structures. Now we state the result which is already implicitly contained
in the constructions of the previous sections \\
\emph{under }(\emph{$A$})\emph{, ($B$), ($C$) the quantization
corresponds to the change of a smoothing of $\mathbb{R}^{4}$ (from
an exotic to the standard one). }\\
This is what we call the 4-exotics approach to (quantum) theories
of physics. It bears some common features with the topos approach
to \emph{theories of physics (TA)}. Namely, in TA one tries to reformulate
quantum theory internally in some topos such that the results resembles
rather classical, though logically intuitionistic, theory. So one
can say that in TA\emph{}\\
\emph{quantization corresponds to the change of a category (from a
topos to ${\rm Set}$) and the corresponding change of logic.}\\
Certainly, as we already mentioned, our 4-exotics approach up to now,
is limited to dimension 4 and to specific theories. The appearance
of the factor $I\! I\! I_{1}$ is a common feature for \emph{all }small
exotic $\mathbb{R}^{4}$'s (as being classified by the codimension-1
foliations of $S^{3}$), so we do not vary between these exotics at
this stage. 

Let us present some more details of the approaches. The main technical
ingredients of the 4-exotics approach of the quantization (related
to a small exotic $\mathbb{R}^{4}$) was already developed in the
previous sections. Let us focus now on TA. 

The topos approach to theories of physics as reformulated after D\"oring
and Isham \cite{Isham:08,Isham:08a,Isham:08b,Isham:08c} by Caspers,
Heunen, Landsman and Spitters \cite{Landsman:07,Landsman:09,Landsman:09a}
is based on the algebraic formulation of QM i.e. given by a $C^{\star}$
algebra, $A$, whose self-adjoint elements represent the algebra of
observables of a quantum theory. Given such an algebra $A$ (noncommutative)
one tries to build the spectrum $\Sigma_{A}$ of it by the analogy
with the Gelfand-Naimark spectrum $X$ existing for a commutative
$C^{\star}$ algebra. This means that one tries to model the algebra
$A$ by the 'isomorphic' to it object of 'functions' $\underline{f}:\Sigma_{A}\to\underline{R}$
where $\underline{R}$ is the suitable counterpart to the set of real
numbers. The above goal can not be achieved for the 'true' (i.e. in
the mathematical sense based on classical logic) real numbers, true
topological compact space $\Sigma_{A}$ and true continuous functions
on $\Sigma_{A}$. The big achievement of the topos approach to quantum
theories is to solve this problem for some categories in mathematics,
namely topoi. Thus $\Sigma_{A}$, $\underline{R}$ are objects in
a topos $\tau_{A}$ uniquely determined by the algebra $A$ and $\underline{f}$'s
are morphisms in $\tau_{A}$. The 'isomorphisms' is now understood
as internal in the topos. The topos $\tau_{A}$ which corresponds
to the noncommutative algebra $A$ and makes it a commutative internal
one is the topos of covariant functors on $C(A)$ where $C(A)$ is
the category of commutative subalgebras of $A$ and their homomorphisms.
This whole procedure requires however the internal logic and set theory
of topoi which is not classical but rather \emph{intuitionistic}.
In general, the usual mathematical constructions can be formulated
in topoi provided they make no use of the set theoretical axiom of
choice or the logical rule of the excluded middle. 

The existence of the internal spectrum $\Sigma_{A}$ for a noncommutative
$C^{\star}$ algebra $A$ was established in a series of papers by
Banashewski and Mulvey \cite{Mulvey1,Mulvey2,Mulvey3}. $\Sigma_{A}$is
a (completely regular) locale in $\tau_{A}$ which means that it generalizes
ordinary topological spaces towards the pointfree topologies. Such
a locale $\Sigma_{A}$ plays a role of quantum phase space of a system
under consideration and determines the space of its subobjects. Similarly
as closed linear subspaces of a Hilbert space ${\cal H}$ define a
quantum logic (according to the original proposition by Birkhoff and
von Neumann \cite{vNeumann1}) or propositions about the system described
by ${\cal H}$ (i.e. the open subsets of $\Sigma_{A}$) define the
logic of our internal system as well about the algebra $A$. The space
of open subsets of $\Sigma_{A}$ is a Heyting algebra. The elementary
fact of the theory of Heyting algebras is that these are distributive
in opposition to the nondistributive lattice of projections on closed
linear subspaces of ${\cal H}$. Thus the logic of previously nondistributive
quantum lattice of projections was weakened to the distributive Heyting
algebras, i.e. to the intuitionistic logic of topoi. The noncommutativity
of the original $C^{\star}$ algebra was inverted to the internal
commutative $C^{\star}$ algebra in $\tau_{A}$. The price to pay
is, however, the essential use of intuitionistic logic of topoi. Next
people try to develop all ingredients of this internal intuitionistic
theory as if it were a classical theory and we will present these
below following \cite{Landsman:07}. 

The topos techniques are also applicable in our case of the factor
\emph{${\rm III}_{1}$} algebra and the Banach algebra of sections,
$\Gamma{\cal K}_{P}$ as discussed in Secs. \ref{sub:Twisted-K-theory-algebraic},
\ref{sub:Exotic-and-noncommutative algebra}, when extended to a $C^{\star}$
algebra with unity. Given the factor $I\! I\! I_{1}$, say $N$, we
can follow the topos procedure:

\begin{enumerate}
\item We build the category ${\cal C}(N)$ whose objects are unital commutative
subalgebras, $C\subset N$, (this set of subalgebras is partially
ordered by the inclusion), and morphisms between 2 objects, $C,D\subset N$,
i.e. $Hom_{{\cal C}(N)}(C,D)$, contains a single arrow, when $C\subseteq D$,
and $\emptyset$ (no arrow) otherwise. ${\cal C}(N)$ is usually considered
as the category of classical windows of the noncommutative (quantum)
algebra of observables describing a quantum system.
\item It is known that for any category $K$ and the category ${\rm Set}$
of sets and functions, the category of functors $K\to{\rm Set}$ and
natural transformations between these functors, i.e. ${\rm Set}^{K}$,
is a topos. In our case we take the topos \[
\tau_{N}={\rm Set}{}^{{\cal C}(N)}\]

\item The tautological functor $\underline{N}:C\to C$, $C\in{\cal C}(N)$
from ${\cal C}(N)$ to ${\rm Set}$ defines an internal in $\tau_{N}$
\emph{commutative} unital $C^{\star}$ algebra $\underline{N}$. This
algebra is called by Landsman \emph{et.all, }\cite{Landsman:09a}
the Bohrification of $N$ since $N$ is noncommutative whereas $\underline{N}$
is a commutative algebra, and the Bohr doctrine of quantum physics
relays, roughly, on seeing its also experimental results, in a final
stage, via the classical (commutative) way. 
\item Now given a commutative internal $C^{\star}$ algebra in $\tau_{N}$
we can make use of the results of Banashewski and Mulvey regarding
Gelfand spectra \cite{Mulvey3} of the internal in topoi commutative
$C^{\star}$ algebras. Namely, there exists an internal in $\tau_{N}$
spectrum of the algebra $\underline{N}$ which is a completely regular
locale $\Sigma_{N}$. The object of all locale maps from $\Sigma_{N}$
to the object of complex numbers in $\tau_{N}$, $\mathbb{C},$ i.e.
$\tau_{N}(\Sigma_{N},\mathbb{C})$ is a commutative internal $C^{\star}$
algebra which is isomorphic (internally) to $\underline{N}$: \[
\tau_{N}(\Sigma_{N},\mathbb{C})\simeq_{\tau_{N}}\,\underline{N}\]

\end{enumerate}
Now we can collect physical counterparts of the above construction:

\begin{enumerate}
\item The quantum phase space of fields system given by the algebra of local
observables $N$ (the factor $I\! I\! I_{1}$) is the above locale
$\Sigma_{N}$ (internal in $\tau_{N}$).
\item The internal open subsets of $\Sigma_{N}$ correspond to the arrows
in $\tau_{N}$, $1\to{\cal O}(\Sigma_{N})$, i.e. a frame corresponding
to the locale $\Sigma_{N}$, where $1$ is the terminal object in
$\tau_{N}$ and ${\cal O}(\Sigma_{N})$ is the lattice of (open) subobjects
of $\Sigma_{N}$. This lattice of open subsets represents the quantum
projections or propositions about the quantum system described by
the algebra $N$. However, this lattice is a distributive Heyting
algebra rather than the external non-distributive von Neumann lattice
of projections. 
\item Each observable $o\in N$ determines a map between locales $\delta(o):\Sigma_{N}\to\Re$
where $\Re$ is an interval-domain object in $\tau_{N}$ which, for
the task of the correct interpreting the quantum theory observables
via internal 'functions', plays the role of internal real numbers.
It is worth mentioning that $\Re$ is not an object of real numbers
in the topos $\tau_{N}$. It is rather suitably smeared version of
the internal reals.
\item Given an open subset-interval of the external real numbers, $\Delta\subseteq\mathbb{R}$,
we have the corresponding internal open subset of $\Re$ hence the
arrow in $\tau_{N}$, $\Delta:1\to{\cal {\cal O}}(\Re)$. Besides,
the map $\delta(o):\Sigma_{N}\to\Re$ as above gives rise to the map
of the frames: $\delta^{-1}(o):{\cal O}(\Re)\to{\cal O}(\Sigma_{N})$.
The composition of both, i.e. $1\overset{\Delta}{\to}{\cal O}(\Re)\overset{\delta^{-1}}{\to}{\cal O}(\Sigma_{N})$,
gives the proposition $[o\in\Delta]:1\to{\cal O}(\Sigma_{N})$.
\item In the Birkhof and von Neumann lattice of projections ${\cal L}({\cal H})$
on the closed linear subspaces $[o\in\Delta]$'s these are images
of the spectral projections $E(\Delta){\cal H}$ of $o$ where $o$
is an observable. The pairing of a state $\rho$ and a proposition
$o\in\Delta$ is given by the Born rule (for pure states) $<\rho,o\in\Delta>$
which is a number in $[0,\:1]$. In the topos interpretation of QM
we have the corresponding pairing given by \[
<o\in\Delta,\rho>=1\overset{[o\in\Delta]}{\to}{\cal O}(\Sigma_{N})\overset{\chi_{\rho}}{\to}\Omega_{N}\]
 where $\chi_{\rho}$ is the characteristic function of the state
$\rho$ as an subobject of ${\cal O}(\Sigma_{N})$ and $\Omega_{N}$
is the subobject classifier of $\tau_{N}$. 
\end{enumerate}
The above topos procedure applies also to the factor $I\! I\! I_{1}$
algebra and via the uniqueness of this factor to an algebraic relativistic
QFT (RAQFT) defined on Minkowskian space $M^{n}$. To deal with the
algebraic and local structure of this QFT one notices that an AQFT
can be defined as a functor ${\rm AQFT}:{\cal O}(M^{n})\to{\rm CSTAR}$
\cite{Landsman:07} from the category of open subobjects in the Minkowski
space and their inclusions to the category of $C^{\star}$ algebras
and their morphisms. From this more general perspective one can take
as a topos for this AQFT $\tau_{{\rm AQFT}}={\rm Set}^{{\cal O}(M^{n})}$and
a single internal commutative $C^{\star}$ algebra. As a result of
the topos approach to AQFT on the 4-Minkowski space we obtain an intuitionistic
classical-like internal theory representing the local algebraic QFT. 

Let us observe that the 4-exotic approach to the factor $I\! I\! I_{1}$
and RAQFT can give a somewhat similar result. The factor $I\! I\! I_{1}$
of a relativistic algebraic QFT defined on $M^{4}$ can be obtained
by the deformation of the Poisson algebra of some classical theory
as was presented in Secs. \ref{sub:From-exotic-smoothness}, \ref{sub:The-observable-algebra}.
However, this whole procedure (via our conjectures at the beginning
of this section) corresponds to the change of exotic smooth structure
on $\mathbb{R}^{4}$, to the standard smooth $\mathbb{R}^{4}$ (ew.
with Minkowskian metric hence $M^{4}$). The point is that one can
in principle formulate a classical field theory on an exotic $\mathbb{R}^{4}$
and this should correspond to some quantized AQFT on $M^{4}$. Thus
the task of quantization of some AQFT on $M^{4}$ and the reverse
task to obtain a classical theory from a quantum relativistic AQFT,
can be formulated in the 4-exotic paradigm as:

\emph{Let CFT be a classical field theory on $\mathbb{R}^{4}$ or
$M^{4}$ which by assumption (C) determines a classical theory with
the algebra of observables equal to the Poisson algebra $(X(S,SL(2,\mathbb{C}),\left\{ \,,\,\right\} )$
of complex functions on $X(S,SL(2,\mathbb{C}))$. The quantization
of the Poisson algebra can be obtained as a CFT on an exotic $\mathbb{R}^{4}$.
Then changing the smooth structure to the standard $\mathbb{R}^{4}$
gives the factor $I\! I\! I_{1}$ algebra which is the algebra of
quantum observables of the quantized theory. This quantum theory is
equivalent by assumption (B) to some RAQFT on $M^{4}$. The equivalence
can be understood as a duality of theories. Conversely, a quantized
RAQFT on the standard smooth $\mathbb{R}^{4}$ with Minkowski metric,
is derived from a classical field theory on certain exotic smooth
$\mathbb{R}^{4}$. }

Hence the classical theory on the exotic $\mathbb{R}^{4}$ is dual
in the above sense to some quantized AQFT. A relativistic AQFT on
standard smooth $\mathbb{R}^{4}$ with Minkowski metric, becomes a
classical field theory with the Poisson algebra when formulated on
certain small exotic $\mathbb{R}^{4}$. In this case the quantization
is performed via the change of the smooth structure of the exotic
$\mathbb{R}^{4}$ to the standard one. This is analogous to the change
of the logic from the intuitionistic logic of topoi to the classical
logic of Set in the topos approach, which corresponds to quantization.

We do not have explicit descriptions of a RAQFT on an exotic $\mathbb{R}^{4}$
or even classical field theory on it since we do not have an exotic
metric nor the global exotic smooth structures glued from local coordinate
patches. However the local descriptions are completely equivalent
with the standard case. Moreover, we have the relation between algebras
of observables and exotic smoothings of $\mathbb{R}^{4}$ worked out
in previous subsections. From that algebraic point of view all we
need is the knowledge about the existence of theories which have the
quantum algebra of observables spanned on the factor $I\! I\! I_{1}$
and the classical algebra spanned on a Poisson algebra. 

A natural question one can ask is the relation of the (intuitionistic)
classical theories obtained from the topos approach and via 4-exotics,
provided one starts with the same quantum RAQFT %
\footnote{The connection of topos theory and intuitionism with exotic $\mathbb{R}^{4}$'s,
though from the model-theoretic perspective, was already proposed
in \cite{Krol:04a,Krol:2005}.%
}. Probably this requires more thorough understanding of both approaches
with a well recognized relation between internal intuitionistic and
external theories and their physical contents. Let us just make a
few comments at this point. 

\begin{itemize}
\item Toposes serve as a generalization of topological spaces. This gives
rise to the intuitionistic driven translation of the noncommutative
$C^{\star}$ algebras to commutative ones, though internal in toposes. 
\item In the 4-exotics approach to quantum physics we rather deal with generalized
geometries as given by gerbes on manifolds and so called $B$-fields
as the extension of (pseudo-) Riemannian geometry of manifolds \cite{AsselmeyerKrol2009,AsselmeyerKrol2009a}. 
\item The 4-exotics approach is essentially 4-dimensional. The factor $I\! I\! I_{1}$
von Neumann algebra is unique. When one wants to vary different exotic
$\mathbb{R}^{4}$'s in this approach, the net of algebras suitably
embedded into each other should be probably considered. However even
in dimension 2 the description of such nets appears as quite nontrivial
task \cite{Longo:2004}. 
\end{itemize}
The topos approach is universally valid for any unital $C^{\star}$
algebra and intuitionism is an essential technical and conceptual
ingredient of this. However, one does not have direct relation of
classical (internal in toposes) and external quantum observables.
From the other side, the 4-exotics approach is certainly up to now
bounded to the limiting case of theories, namely those which have
the factor $I\! I\! I_{1}$ as the subalgebra of the observables algebra,
and works exclusively in dimension 4. However, dimension 4 is of exceptional
importance for physics and the factor $I\! I\! I_{1}$ is also quite
essential for relativistic local field theories. All constructions
are performed in classical mathematics without necessity to refer
to intuitionism. We hope to present soon the more detailed analysis
of the 4-exotics approach with some valid for physics applications.

\section{Conclusions}

In this paper we presented a variety of relations between codimension-1
foliations of the 3-sphere $S^{3}$ and noncommutative algebras. By
using the results of our previous paper \cite{AsselmeyerKrol2009},
we obtain a relation between (small) exotic smoothness of the $\mathbb{R}^{4}$
and noncommutativity via the noncommutative leaf space of the foliation
and the Casson handle. Thus we get our main result of this paper:
\\
\emph{The Casson handle carries the structure of a noncommutative
space determined by a factor $I\! I_{1}$ algebra which is related
to the skein algebra of the disk with marked points and to the leaf
space of the horocycle foliation.} \\
Thus we have obtained a direct link between noncommutative spaces
and exotic 4-manifolds which can be used to get a direct relation
to quantum field theory. One of the central elements in the algebraic
quantum field theory is the Tomita-Takesaki theory leading to the
$I\! I\! I_{1}$ factor as vacuum sector \cite{Borchers2000}. As
a possible candidate one has loop quantum gravity with an unique diffeomorphism-invariant
vacuum state \cite{Thiemann2006}.

The attentive reader came across with the fact that there are one
hyperfinite $I\! I\! I_{1}$factor but many Casson handles. We will
discuss this possible contradiction now. In the subsection \ref{sub:factor-III-case}
we presented the relation $I\! I\! I_{1}=R_{\infty}=R_{\lambda_{1}}\otimes R_{\lambda_{2}}$
for the Araki-Woods factor for all $\lambda_{1},\lambda_{2}$ with
$\lambda_{1}/\lambda_{2}\notin\mathbb{Q}$ to the Powers factors $I\! I\! I_{\lambda}=R_{\lambda}$.
Thus we have a continuum of $I\! I\! I_{1}$ factors parametrized
by irrational numbers $\lambda_{1}/\lambda_{2}\notin\mathbb{Q}$.
It is known by Freedmans work \cite{Fre:82} that all Casson handles
can be parametrized by a dual tree or by the Cantor continuum. Thus
every particular Casson handle is given by a real (irrational) number.
All the corresponding factors are equivalent to $I\! I\! I_{1}$and
we need finer invariants to distinguish them from each other. One
possible invariant is the flow of weights introduced by Connes \cite{Connes94}
and related to the Godbillon-Vey invariant. That closes the circle
to the elements of $H^{3}(S^{3},\mathbb{R})$ and we obtain the description
of the whole group $H^{3}(S^{3},\mathbb{R})$ in contrast to the integer
case in \cite{AsselmeyerKrol2009a}. 

In the retrospective, we have many close and unexpected relations
between exotic smooth $\mathbb{R}^{4}$ and string theory especially
WZW models and D-brane charges. With this paper we have also included
quantum aspects in our model. But string or M theory is usually formulated
in 10 or 11 dimensions whereas our approach is based on 4 dimensions.
Usually one compactifies partly the 10- or 11-dimensional space to
get the desired 4-dimensional spacetime. In our forthcoming paper
we will show that the structure of the compactified space as 6-dimensional
Calabi-Yau manifold or as 7-dimensional $G_{2}$ manifold can be also
induced by the smoothnes structure of the spacetime.

\appendix

\section{\label{sec:Non-cobordant-foliationsS3}Non-cobordant foliations of
$S^{3}$ detected by the Godbillon-Vey class}

In \cite{Thu:72}, Thurston constructed a foliation of the 3-sphere
$S^{3}$ depending on a polygon $P$ in the hyperbolic plane $\mathbb{H}^{2}$
so that two foliations are non-cobordant if the corresponding polygons
have different areas. We will present this construction now.

Consider the hyperbolic plane $\mathbb{H}^{2}$ and its unit tangent
bundle $T_{1}\mathbb{H}^{2}$ , i.e the tangent bundle $T\mathbb{H}^{2}$
where every vector in the fiber has norm $1$. Thus the bundle $T_{1}\mathbb{H}^{2}$
is a $S^{1}$-bundle over $\mathbb{H}^{2}$. There is a foliation
$\mathcal{F}$ of $T_{1}\mathbb{H}^{2}$ invariant under the isometries
of $\mathbb{H}^{2}$ which is induced by bundle structure and by a
family of parallel geodesics on $\mathbb{H}^{2}$. The foliation $\mathcal{F}$
is transverse to the fibers of $T_{1}\mathbb{H}^{2}$. Let $P$ be
any convex polygon in $\mathbb{H}^{2}$. We will construct a foliation
$\mathcal{F}_{P}$ of the three-sphere $S^{3}$ depending on $P$.
Let the sides of $P$ be labeled $s_{1},\ldots,s_{k}$ and let the
angles have magnitudes $\alpha_{1},\ldots,\alpha_{k}$. Let $Q$ be
the closed region bounded by $P\cup P'$, where $P'$ is the reflection
of $P$ through $s_{1}$. Let $Q_{\epsilon}$, be $Q$ minus an open
$\epsilon$-disk about each vertex. If $\pi:T_{1}\mathbb{H}^{2}\to\mathbb{H}^{2}$
is the projection of the bundle $T_{1}\mathbb{H}^{2}$, then $\pi^{-1}(Q)$
is a solid torus $Q\times S^{1}$(with edges) with foliation $\mathcal{F}_{1}$
induced from $\mathcal{F}$. For each $i$, there is an unique orientation-preserving
isometry of $\mathbb{H}^{2}$, denoted $I_{i}$, which matches $s_{i}$
point-for-point with its reflected image $s'_{i}$. We glue the cylinder
$\pi^{-1}(s_{i}\cap Q_{\epsilon})$ to the cylinder $\pi^{-1}(s'_{i}\cap Q_{\epsilon})$
by the differential $dI_{i}$ for each $i>1$, to obtain a manifold
$M=(S^{2}\setminus\left\{ \mbox{\mbox{k} punctures}\right\} )\times S^{1}$,
and a (glued) foliation $\mathcal{F}_{2}$, induced from $\mathcal{F}_{1}$.
To get a complete $S^{3}$, we have to glue-in $k$ solid tori for
the $k$ $S^{1}\times\mbox{punctures}.$ Now we choose a linear foliation
of the solid torus with slope $\alpha_{k}/\pi$ (Reeb foliation).
Finally we obtain a smooth codimension-1 foliation $\mathcal{F}_{P}$
of the 3-sphere $S^{3}$ depending on the polygon $P$.

Now we consider two codimension-1 foliations $\mathcal{F}_{1},\mathcal{F}_{2}$
depending on the convex polygons $P_{1}$ and $P_{2}$ in $\mathbb{H}^{2}$.
As mentioned above, these foliations $\mathcal{F}_{1},\mathcal{F}_{2}$
are defined by two one-forms $\omega_{1}$ and $\omega_{2}$ with
$d\omega_{a}\wedge\omega_{a}=0$ and $a=0,1$. Now we define the one-forms
$\theta_{a}$ as the solution of the equation\[
d\omega_{a}=-\theta_{a}\wedge\omega_{a}\]
and consider the closed 3-form\begin{equation}
\Gamma_{\mathcal{F}_{a}}=\theta_{a}\wedge d\theta_{a}\label{eq:Godbillon-Vey-class}\end{equation}
 associated to the foliation $\mathcal{F}_{a}$. As discovered by
Godbillon and Vey \cite{GodVey:71}, $\Gamma_{\mathcal{F}}$ depends
only on the foliation $\mathcal{F}$ and not on the realization via
$\omega,\theta$. Thus $\Gamma_{\mathcal{F}}$, the \emph{Godbillon-Vey
class}, is an invariant of the foliation. Let $\mathcal{F}_{1}$ and
$\mathcal{F}_{2}$ be two cobordant foliations then $\Gamma_{\mathcal{F}_{1}}=\Gamma_{\mathcal{F}_{2}}$.
In case of the polygon-dependent foliations $\mathcal{F}_{1},\mathcal{F}_{2}$,
Thurston \cite{Thu:72} obtains\[
\Gamma_{\mathcal{F}_{a}}=vol(\pi^{-1}(Q))=4\pi\cdot Area(P_{a})\]
and thus

\begin{itemize}
\item $\mathcal{F}_{1}$ is cobordant to $\mathcal{F}_{2}$ $\Longrightarrow$$Area(P_{1})=Area(P_{2})$
\item $\mathcal{F}_{1}$ and $\mathcal{F}_{2}$ are non-cobordant $\Longleftrightarrow$$Area(P_{1})\not=Area(P_{2})$
\end{itemize}
We note that $Area(P)=(k-2)\pi-\sum_{k}\alpha_{k}$. The Godbillon-Vey
class is an element of the deRham cohomology $H^{3}(S^{3},\mathbb{R})$.
Furthermore we remark that the classification is not complete. Thurston
constructed only a surjective homomorphism from the group of cobordism
classes of foliation of $S^{3}$ into the real numbers $\mathbb{R}$.

\section*{Acknowledgment}

T.A. wants to thank C.H. Brans and H. Ros\'e for numerous discussions
over the years about the relation of exotic smoothness to physics.
J.K. benefited much from the explanations given to him by Robert Gompf
regarding 4-smoothness several years ago, and discussions with Jan
Sladkowski.


\end{document}